\documentclass[twocolumn,showpacs,amsmath,amssymb,aps,prd,floats]{revtex4-1}
\usepackage{graphicx}% Include figure files
\usepackage[switch,pagewise,running]{lineno}
\usepackage{dcolumn}% Align table columns on decimal point
\usepackage{bm}% bold math
\usepackage{epsfig}
\usepackage{xspace}
\usepackage{hyperref}
\hypersetup{colorlinks,linkcolor=red,citecolor=blue,urlcolor=blue}  
\usepackage{color,xcolor}
\usepackage{verbatim}
\usepackage{float}
\usepackage{slashed}
\usepackage{rotating}
\usepackage[normalem]{ulem} 
\usepackage[caption=false]{subfig}
\usepackage{verbatim}
\usepackage{threeparttable}

\def\apjl{The Astrophysical Journal Letter}
\def\apjs{The Astrophysical Journal Supplement}

\def\aap{Astronomy and Astrophysics}

\newcommand{\erg}{${\rm erg \ s^{-1}}$ }

\def\ltsima{$\; \buildrel < \over \sim \;$}
\def\simlt{\lower.5ex\hbox{\ltsima}}
\def\gtsima{$\; \buildrel > \over \sim \;$}
\def\simgt{\lower.5ex\hbox{\gtsima}}
\newcommand{\msun}{{\rm\,M$_\odot$}}

\newcommand{\srcs}{{\rm\,J1513+3111}}
\newcommand{\src}{{\rm\,J1513+3111 }}

\begin{document}

\title{
%A high-energy neutrino associated with radio emission from TDE candidate SDSS J151345.75+311125.2}
Dust-obscured radio-emitting tidal disruption event coincident with a high-energy neutrino event}
\author{Tianyao Zhou$^{1}$}
\author{Xinwen Shu$^{1}$}\email{xwshu@ahnu.edu.cn}
\author{Guobin Mou$^{2}$}\email{gbmou@njnu.edu.cn}
\author{Lei Yang$^{1}$} %ahnuyl@ahnu.edu.cn
\author{Luming Sun$^{1}$} 
\author{Fangkun Peng$^{1}$} %
\author{Fabao Zhang$^{1}$}
\author{Hucheng Ding$^{1}$}
\author{Ning Jiang$^{3}$}
\author{Tinggui Wang$^{3}$}
\author{Yogesh Chandola$^{4, 5}$} %yogesh.chandola@gmail.com
\author{Daizhong Liu$^{5}$} %dzliu@pmo.ac.cn 
\author{Liming Dou$^{6}$} %doulm@gzhu.edu.cn
\author{Yibo Wang$^{3}$} %wybustc@mail.ustc.edu.cn
\author{Jianguo Wang$^{7}$} %wangjg@ynao.ac.cn
\author{Zhongzu Wu$^{8}$} %gzu_zzwu@163.com
\author{Chenwei Yang$^{9}$} %yangchenwei@pric.org.cn

\affiliation{$^{1}$Department of Physics, Anhui Normal University, Wuhu 241002, China
}
\affiliation{$^{2}$Department of Physics and Institute of Theoretical Physics, Nanjing Normal University, Nanjing 210023, China
}

\affiliation{$^{3}$Department of Astronomy, University of Science and Technology of China, Hefei, Anhui 230026, China
}
\affiliation{$^{4}$Indian Institute of Astrophysics (IIA), 2nd Block, Koramangala, Bengaluru, 560034, Karnataka, India.
}

\affiliation{$^{5}$Purple Mountain Observatory, Chinese Academy of Sciences, Nanjing 210023, China
}

\affiliation{$^{6}$Department of Astronomy, Guangzhou University, Guangzhou 510006, China} 

\affiliation{$^{7}$Yunnan Observatories, Chinese Academy of Sciences, Kunming 650011, China
}

\affiliation{$^{8}$College of Physics, Guizhou University, Guiyang 550025, China}

\affiliation{$^{9}$Polar Research Institute of China, 451 Jinqiao Road, Shanghai 200136, China
}

\begin{abstract}
Despite the growing number of high-energy neutrinos (TeV–PeV) detected by IceCube, their astrophysical origins remain largely unidentified. Recent observations have linked a few tidal disruption events (TDEs) to the production of high-energy neutrino emission, all of which display dust-reprocessed infrared flares indicating a dust and gas-rich environment. %During 
By cross-matching the neutrino events and a sample of mid-infrared outbursts in nearby galaxies with transient radio flares, we uncover an optically obscured TDE candidate, SDSS J151345.75+311125.2, which shows both spatial and temporal coincidence with the sub-PeV neutrino event IC170514B. 
Using a standard equipartition analysis of the synchrotron spectral evolution spanning 605 days post mid-infrared discovery, we find a little evolution in the radio-emitting region, 
%significant  the outflow powering the radio emission is significantly decelerated, 
%while the minimum kinetic energy decreases \textbf{$E_{\rm K}\approx4.4\times10^{49}$ erg to $E_{\rm K}\approx2.9\times10^{49}$ erg.}
with a kinetic energy up to $10^{51}$ erg, depending on the outflow geometry and shock acceleration efficiency assumed.
%The size of radio-emitting region can be further constrained by the 
High-resolution EVN imaging reveals a compact radio emission that is unresolved at a scale of $<$2.1 pc, 
with a brightness temperature of $T_b>5\times10^6$ K, suggesting that the observed late-time radio emission might originate from the interaction between a decelerating outflow and a dense circumnuclear medium. 
If the association is genuine, the neutrino production is possibly related to the acceleration of protons through $pp$ collisions during the outflow expanding process,    
implying that the outflow-cloud interaction could provide a physical site with high density environment for producing the sub-PeV neutrinos. %, {\bf though a high cloud density is required}.  
%This work implies that the hardronic process in the outflow-cloud interaction may play a vital role in explaining the cosmic neutrino flux. 
Such a scenario can be tested with future identifications of radio transients 
coincident with high-energy neutrinos.

%We conducted multi-epoch radio observations of J1513+3111 using the VLA, GMRT, and EVN. These observations revealed a rise in radio flux density coincided with the arrival time of the neutrino event, supporting a potential physical connection. We model the evolution of the radio spectral energy distribution via synchrotron emission from relativistic electrons, considering a scenario where a TDE outflow interacts with the circumnuclear medium. We derived the kinetic energy of outflow and the radius of the radio emission (potential potential position of particles acceleration) 1-3 years after the neutrino detection time to be $\sim 10^{50} \ {\rm erg}$ and $\sim 10^{17} \ {\rm cm}$, respectively. High-energy protons accelerated in this process could produce neutrinos through $pp$ collisions.
\end{abstract}

\maketitle

\section{Introduction}
High-energy ($>$100 TeV) neutrinos is a smoking-gun signature of hadronic reaction and therefore traces the origin, acceleration, and propagation processes of high-energy cosmic rays. In 2013, the IceCube Neutrino Observatory detected a diffuse astrophysical neutrino flux above 30 TeV %at the South Pole 
for the first time, which opens a new era for multi-messenger astronomy and astroparticle physics \cite{IceCube2013,2014PhRvL.113j1101A}. During continuous operation of the IceCube detector for more than 10 years, the increasing number of high-energy neutrino events is collected (e.g., \cite{2022ApJ...928...50A,Abbasi2023}). Despite extensive searches for neutrino excesses over the background across multiple astrophysical source classes (e.g., \cite{2017ApJ...835..151A,2017ApJ...835..269P,2022ApJ...926...59A,2022PhRvD.106b2005A,2023ApJ...949L..12A,2025arXiv250309426L}), IceCube has yet to reach conclusive evidence of identifying a dominant source. %electromagnetic counterpart. 

%source.

To date, only a few candidates of high-energy neutrino emission were tentatively confirmed with strong evidence by the IceCube Collaboration, including the blazar TXS 0506+056 \cite{IceCube2018}, the nearby Seyfert galaxy NGC 1068 \cite{IceCube2022}, and the Galactic plane \cite{2023Sci...380.1338I}. Recently, an optical tidal disruption event (TDE) AT2019dsg discovered by the Zwicky Transient Facility (ZTF) was found to be in spatial and temporal coincidence with the high-energy neutrino IC191001A \cite{Stein2021}.  %which is the first likely association between the IceCube neutrino and TDE. 
%This suggests that TDEs could be n
%, rare accretion flares that occur when 
%The distinctive properties of AT2019dsg include a prominent thermal spectrum in the optical-ultraviolet (OUV) range, delayed bright dust echo emission in the infrared, as well as high X-ray luminosity and long-term radio emission. 
Almost at the same time, 
%two TDE candidates, AT2019fdr and AT2019aalc with similar optical and infrared emission features, were uncovered as a promising neutrino source by multi-messenger analysis \cite{Reusch2022,Velzen2024}. 
another TDE candidate in ZTF, AT2019fdr, was uncovered as a promising neutrino source, though occurring in 
an active galactic nucleus \cite[AGN,][]{Reusch2022}. 
The distinctive property of the two neutrino-coincident ZTF sources is the delayed infrared (IR) flare, likely originating from the reprocessed dust emission of central outburst \cite{Jiang2021b}. 
These findings motivate a systematic search for neutrino emission from TDEs with similar dust echoes, 
resulting in three more events, including two dust-obscured TDEs that are faint in optical \cite{Jiang2023, Velzen2024}. 
%This established that 
%both observational and theoretical perspectives. Later, two candidate obscured TDEs were reported to correlate with two high-energy track-like neutrinos by matching the sample of mid-infrared outbursts in nearby galaxies (MIRONG) with gold-type events (signalness at least $\sim 50\%$ to be of astrophysical origin) \cite{Jiang2021b,Jiang2023}. 
When the match criteria are set loosely, for example TDEs located slightly beyond the 90\% error region of the corresponding neutrinos or neutrino events with a lower signalness are considered, another two TDE candidates, AT2021lwx \cite{Yuan2024} and ATLAS17jrp \cite{Li2024} are pointed out to coincide potentially with the high-energy neutrino IC220405B and a candidate neutrino flare, respectively.

%TDEs are energetic high-energy transients when a star is destroyed by a supermassive black hole (SMBH) within its tidal radius \cite{1988Natur.333..523R}. Part of the stellar debris remains bound and is accreted by SMBH, which can result in a luminous emission from the radio to gamma-ray bands on timescales of months to years. 
While it appears that the efficiency of high-energy neutrino production in 
TDEs and related accretion flares from supermassive black holes (SMBHs) is high 
compared to non-flaring AGN \cite{Velzen2024}, the detailed physical mechanisms 
are far from clear. 
%TDEs that occur when an unlucky star approaches close enough to a supermassive black hole
%(SMBH) and gets ripped apart by the tidal force, are promising new class astrophysical 
Theoretically, it has been speculated that a relativistic jet has the advantage of providing the sufficient power for particle acceleration, early study has suggested that high-energy neutrinos could be produced via the $p\gamma$ interactions in the jet environment of TDEs \cite{2011PhRvD..84h1301W}. The jet model was carefully studied in subsequent works \citep{Wang2016,2017MNRAS.469.1354D,2017PhRvD..95l3001L,2017ApJ...838....3S,Biehl2018,Guepin2018}. However, no convincing jet activity has been detected in the neutrino-coincident TDEs. %majority of TDEs, particularly after 
Since the detection of AT2019dsg–IC191001A association, other different scenarios are further proposed to produce neutrino emission and multi-band electromagnetic radiation. These involve accretion disks and tidal stream interactions \cite{2019ApJ...886..114H}, hot coronae \cite{2020ApJ...902..108M, Winter2021}, off-axis or choked jetted models \cite{2020PhRvD.102h3028L,2023ApJ...954...17Z,2024MNRAS.534.1528M, Sato2024}, subrelativistic wind \cite{Reusch2022,Yuan2024b}, outflow-cloud interactions \cite{Mou2021, Wu2022}, and a model without a specific acceleration sites \cite{Winter2023, Yuan2023}. 

The shared characteristic of all the seven high-energy neutrino-associated events (candidates) 
mentioned above is the elevated infrared emission, suggesting a dust and gas-rich environment in these TDEs. %When we search for the candidates for IceCube neutrinos, we should pay for a special attention to the radio emission, as  radio observations confirm that particle acceleration is indeed occurring. From 
Both observations and numerical simulations suggest that wide angle, ultrafast outflow is an essential or even ubiquitous component in TDEs (e.g., \cite{2016ApJ...819L..25A,2018MNRAS.474.3593K,2018ApJ...859L..20D,2020MNRAS.492..686L,2019MNRAS.483..565C, Cendes2024}). %Consequently, one would expect to detect the late radio emission due to interaction of outflow with cloud \cite{2024ApJ...971..185C}, as observed in the first three association events (AT2019dsg, AT2019fdr, and AT2019aalc) \cite{Stein2021,2021ApJ...919..127C,Reusch2022,Velzen2024}. 
When the outflow collides with %circumnucealr medium and 
gaseous clouds surrounding TDEs, electrons can be effectively accelerated within a bow shock, which in turn produce non-thermal synchrotron emission at radio wavelengths \cite{Mou2022, Zhuang2025}. Meanwhile, the protons can be accelerated up to several tens PeV therein simultaneously by diffusive shock acceleration processes, in which the high-energy neutrino emission is generated \cite{Wu2022}. 
Interestingly, among the first three neutrino-association TDEs (AT2019dsg, AT2019fdr, and AT2019aalc) \cite{Stein2021,2021ApJ...919..127C,Reusch2022,Velzen2024}, 
%the radio observations revealed that while AT2019aalc displays only marginal variation, the other two sources 
two exhibit variable (transient) radio emission during the period of neutrino arrival.  
Within the framework of the outflow-cloud interaction scenario in TDEs, it is reasonable to infer that we can detect more high-energy neutrino events accompanied by radio and infrared emissions, %like the first TDE AT2019dsg, 
if there are timely and sensitive radio observations. 

In this work, we collect a sample of nuclear accretion flares with both dust echoes and transient radio emission, and investigate their correlation with high-energy neutrinos. Our search uncovers a new obscured radio-emitting TDE candidate 
%SDSS J151345.75+311125.2, 
which coincides with a $\sim 0.2$ PeV neutrino event IC170514B. % when the radio emission and infrared emission enhance.  
%In Section II, the sample and selection are presented. 

\section{Sample and Selection}\label{sample}
We construct a sample that includes both radio and IR flares by searching for radio transients in the sample of mid-infrared outbursts in nearby galaxies \cite[MIRONG,][]{Jiang2021}.  
MIRONG were built by crossmatching the mid-IR outbursts with the low-redshift ($z<0.35$) Sloan Digital Sky Survey (SDSS) spectroscopic galaxies, using the data from Wide-field Infrared Survey Explorer (WISE) \cite{Wright2010} and its new mission the Near-Earth Object WISE (NEOWISE) \cite{Mainzer2014} up to the end of 2018. 
Among a total of 137 sources in the MIRONG sample, 26 were classified as Seyfert \uppercase\expandafter{\romannumeral1} and 23 were classified as Seyfert \uppercase\expandafter{\romannumeral2}. 
In this work, we focus mainly on the sample consisting of TDE candidates, 
%from 2014 to the end of 2018 \cite{Jiang2021b}, %weexcluded 39 were classified as Seyfert galaxies. 
thus exclude 49 Seyfert galaxies from our following analysis. 
This will help to avoid confusion by the AGN emission when discussing the potential association of TDEs with neutrinos, as AGNs could display flux and spectroscopic evolution properties mimicking TDEs \cite{Trakhtenbrot2019, Zabludoff2021}. 
This leaves a sample of 88 MIRONG galaxies.
% and remained 98 sources. 
%The mid-infrared (MIR) light curves were built by the Wide-field Infrared Survey Explorer (WISE) \cite{Wright2010} and its new mission the Near-Earth Object WISE (NEOWISE) \cite{Mainzer2014}. 
For radio observations, on one hand, we used public data based on the Very Large Array Sky Survey (VLASS) \cite{Lacy2020}. VLASS is a multi-epoch S band (2-4 GHz) all-sky survey program, which was initiated in 2017. 
At the time of writing, the complete datasets from the first two epochs and partial data from the ongoing third phase are publicly available. 
On the other hand, we conducted follow-up observations of 35 MIR outbursts in the MIRONG sample using Karl G. Jansky Very Large Array (VLA) at L (1-2 GHz), C (4-8 GHz), X (8-12 GHz) bands. 
%most of which have optical spectroscopy follow-up observations to secure their nuclear transient nature \cite{Wang2022}. 
Our selection of MIR outbursts for VLA observations is based mainly on  those having optical spectroscopic follow-up observations, totally 54 galaxies \cite{Wang2022}. 
In \cite{Wang2022}, we find that 22 out of 54 galaxies show emission-line variations, among which at least 60\% could be interpreted as due to TDEs. %\footnote{\bf The nature of the other 32 objects without clear spectral variations is still mysterious, but could be an interesting class of dust-obscured nuclear flares such as TDEs.}. 
Note that galaxies that were detected in the Faint Images of the Radio Sky at Twenty cm (FIRST) survey are also excluded in our sample for VLA observations.  

A portion of these data has been analyzed and presented in \cite{Dai2020}. The radio transients were selected based on two criteria: (i) non-detections ($3 \sigma$ upper limits) in earlier epochs followed by detections in subsequent epochs; (ii) A flux density increase by a factor of $\ge 1.5$ in at least one radio band. 
%Further details of the selection are provided in \cite{Zhang2022}. 
Further details of the selection and studies of radio evolution properties will be presented elsewhere, which are beyond the scope of current work. 
A final sample of 9 MIR outbursts with transient radio emission (MIR-radio transients) was assembled and will be analyzed in the subsequent sections.

The recently issued IceCube Event Catalog of Alert Tracks (ICECAT-1) \cite{Abbasi2023} contains 275 probable astrophysical neutrino track events detected between May 2011 and December 2020. ICECAT-1 documents the arrival times, reconstructed energies, directional coordinates, spatial uncertainties, and probabilities of astrophysical origin for neutrino events. Neutrino alerts are categorized into gold and bronze channels, corresponding to candidates with 50\% and 30\% probabilities of astrophysical origin, respectively. 
In addition, since 2016 the General Coordinates Network (GCN) reported real-time alerts of high-energy ($>100$ TeV) track-like neutrinos, and updated the positions and errors of previous events in ICECAT-1 when applicable. 
%would be updated through GCN by taking the potential systematic uncertainties of detector into account.

%Since 2016, IceCube has updated the real-time alerts on General Coordinates Network (GCN) platform while detecting high energy ($>$ 100TeV) track-like neutrino events. Due to the more superior angular resolution than cascades, tracks are best suited for use in multi-messenger searched for astrophysical sources. With the improvement of real-time system in 2019, according to neutrino candidates with 50\% and 30\% probability of astrophysical origin respectively, All alerts are classified for gold or bronze channels.  Recently, the IceCube Event Catalog of Alert Tracks (ICECAT-1), containing likely astrophysical neutrino track-like events from 2011 May to the end of 2020 has been presented \cite{Abbasi2023}. For alert directions, the errors on R.A. and Decl. correspond to the 90\% uncertainty likelihood contours with rectangular boundaries. Compared with the real-time alert events released by GCN, the direction and error of neutrino events in ICECAT-1 will be reconstructed, and errors will take into account the potential systematic errors from detector uncertainties. Due to the real-time updating on the alert system, in our work, we selected the gold alerts between 2014 and 2022 June, and we only considered neutrinos with errors less than 200', including a total of 75 events. 

\begin{figure*}
    \centering
    \includegraphics[scale = 0.6]{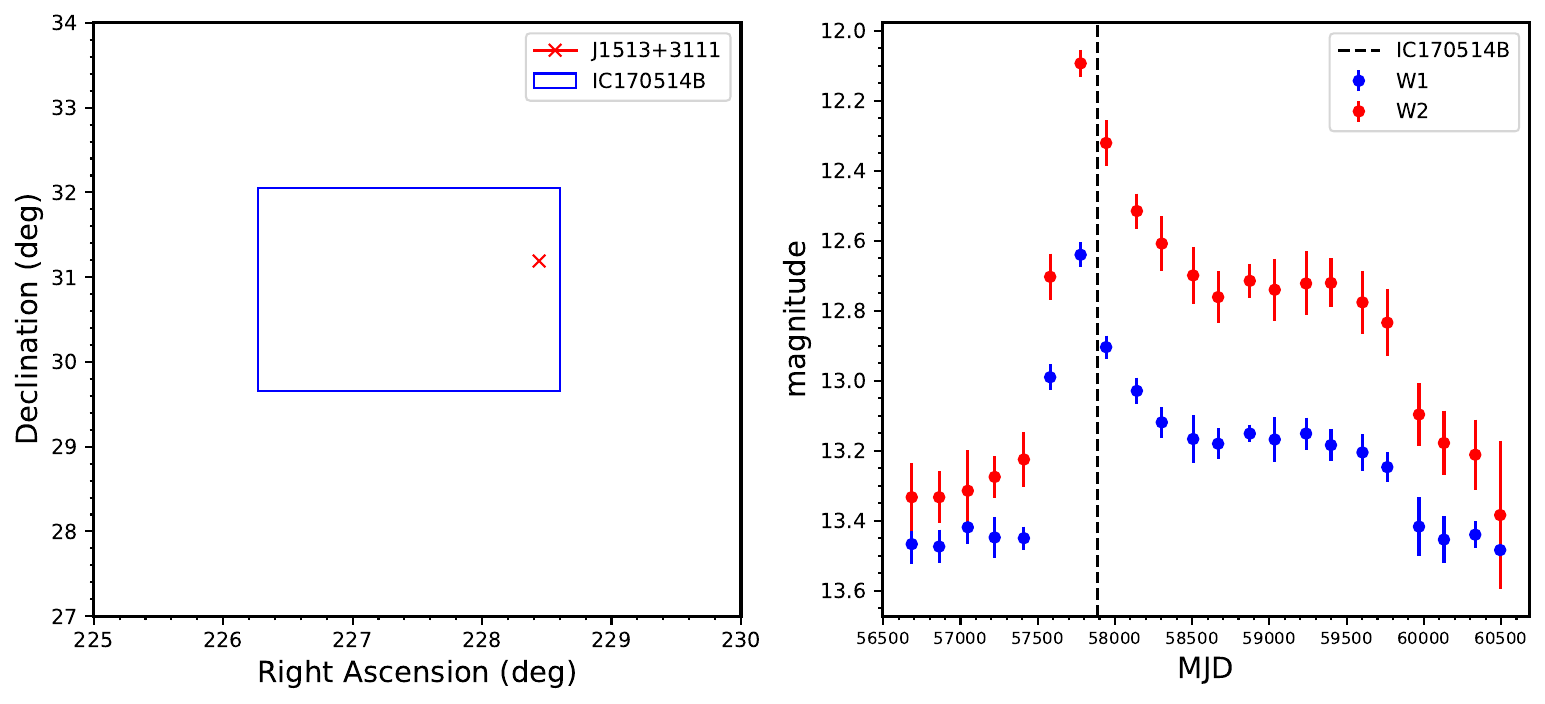}
    \caption{Left panel: Localization of J1513+3111 and IC170514B. The blue rectangular shows the 90\% CL containment region of IC170514B. Right panel: The mid-infrared light curves of J1513+3111 observed by WISE. The vertical dashed line marks the arrival time of matched high-energy neutrino event IC170514B, which delayed the peak of mid-infrared emission by $\sim$109 days.}
    \label{mir}
\end{figure*}

In this work, we selected the gold-type neutrino events from ICECAT-1 (to ensure the best signalness) between 2014 and 2018, which overlaps the period of MIRONG sample's data collection (spanning $\sim$5 years). We considered only neutrinos with positional errors less than 200 arcmin \cite{Jiang2023}, including a total of 44 events. We used the 90\% containment region of gold-type neutrino events as our spatial matching area for identifying correlated MIR-radio transients. The infrared light curves have been constructed at six-month intervals in WISE's W1 (3.4$\mu$m) and W2 (4.6$\mu$m) bands since observations in 2014. 
%{\bf which is the minimum observing cadence of the WISE survey}. 
For the VLASS observations at S-band, %are being taken over a series of three epochs with a 
the cadence is $\sim$32 months over three epochs covering a period from 2017 to 2024, 
while the cadence of our dedicated VLA radio follow-up observations is $\sim$1-2 years on average.  %$$ since 2017. 
Given the better sampling in the MIR light curves, the temporal matching required that the MIR peak flux in at least one band lies within a 0.5-year window before or after the time of neutrino detection. 
Our selection of the 0.5-year matching window corresponds the minimum observing cadence of the WISE survey, and we tested that changing this to a longer time window would not affect the matching results.
Our analysis revealed that the MIR-radio transient SDSS J151345.75+311125.2 (hereafter
J1513+3111) at redshift $z = 0.07181$ is temporally coincident with the high-energy neutrino event IC170514B, 
which was detected by IceCube on May 14, 2017 (MJD = 57887.3) with an energy of $\sim 174$ TeV and a signalness of about 55\%. 
While the arrival of neutrino displays a delay relative to the peak time of mid-infrared emission by $\sim$109 days (Figure \ref{mir}, right), the time delay is less than the minimum observing cadence of the mid-infrared light curve. 
Therefore, we consider that there is a temporal coincidence between J1513+3111 and IC170514B. 
On the other hand, the best-fit position of IC170514B was given to be R.A. = $227^{\circ}.37^{+1^{\circ}.23}_{-1^{\circ}.1}$, Decl. = $30^{\circ}.65^{+1^{\circ}.4}_{-0^{\circ}.99}$ (J2000), with an angular separation of only 1.07$^\circ$ from J1513+3111 (Figure \ref{mir}, left), suggesting that both are possibly spatially coincident as well. Note that no correlation analysis with IceCat-2 \cite{IceCube2025} was performed because the full data release is not yet available, and further correlation studies with IceCat-2 are required to confirm the neutrino coincidence.
%Such a relatively small separation suggests that 
%Note that the neutrino association with J1513+3111 appears to be valid, even considering the ICECAT-2 with improved reconstructions that will be issued by the IceCube Collaboration \cite{IceCube2025}, as the separation is comparable to 
%will issue a new catalog of probable astrophysical neutrino track events \cite{??}, with improved reconstructions. 
%the median of 90\%  containment areas of 1.3 deg$^2$, with the corresponding 1$\sigma$ errors of $0.7-3.7$ deg$^2$. 
%This would not affect the 
%such a small angular separation 

%the angular offset measured 1.2$^\circ$ relative to the best-fitting neutrino position. To find the radio-MIR transients that are spatially and temporally correlated with the observed neutrino events, we crossmatch the transient coordinates with the 90\% error regions of the gold alerts we selected. And in temporal match, due to the WISE visit cadence for half a year higher than VLASS, the time interval between the neutrino arrival and the IR brightest epoch is less than half a year. 

VLASS %has dedicated $\sim 5500$ hours to survey the entire northern sky with decl. $>-40^{\circ}$ in three epochs, covering 
covers a large cumulative area of 33885 deg$^2$. Compared to VLASS, the smaller footprint of MIRONG galaxies implies that their overlapping region is effectively determined by the SDSS coverage area, which is used to calculate the probability of coincidence with neutrino events. The area of the SDSS footprint ($\Omega_{\rm SDSS}$) and the total 90\% containment region of the neutrinos in the IceCube catalog we used ($\Omega_{\rm IceCube}$) are 9376 $\rm deg^{2}$ and 289.58 $\rm deg^{2}$, respectively. We estimated that the density of MIR-radio transients ($\rho_{s}$) is $n_{s}$/$\Omega_{\rm SDSS}$/$T_{s}$=$\rm 1.9\times10^{-4}deg^{-2}yr^{-1}$, where $n_{s}=9$ is the number of MIR-radio transients and $T_{s}=5$ yr is the time period of MIR peak  covered by the MIRONG sample (between 2014 and 2018).  
%’s data collection. Assuming the matching time window is 1 yr, 
Since we matched the MIR-radio transients with neutrinos by requiring their intervals being $\pm0.5$ yr, the matching time window is thus 1 yr. 
In this case, the expectation value ($\lambda$) for the number of random matches is $\rho_{s}\times\Omega_{eff}=0.013$, where $\Omega_{eff}\approx$$\Omega_{\rm IceCube}\times\frac{\Omega_{\rm SDSS}}{4\pi}$ is the effective 90\% containment area of 44 gold neutrino events in the SDSS footprint. Finally, the Poisson probability of observing at least one match can be calculated as P ($X\geq1) = 1-e^{-\lambda}\approx0.013$.
%The SDSS footprint and the total 90\% containment region of the neutrinos in the SDSS footprint are 9376 deg$^2$ and 289.58 deg$^2$, respectively. Following the procedure in \citep{Jiang2023}, we estimated that the density of MIR-radio transients is 9/9376/5=1.9$\times$10$^{-4}$ deg$^{-2}$ yr$^{-1}$. Assuming the matching time window is 1 yr, the expectation value for the number of random matches is $1.9\times10^{-4}\times289.58=0.056$ and the Poisson probability of observing at least one match is thus $P(X\ge1)=0.054$.

We then calculated the significance of the spatial and temporal coincidence of MIR-radio sources with the IceCube track-like neutrinos using simulations. %The observation of 10 years of 
The sensitivity of IceCube is dependent on the declination and the energy of the source (i.e., the spectrum)\cite{Aartsen2020}.
%IceCube data over an observing period of 10 years show that the neutrino flux is sensitive to different declinations \cite{Aartsen2020}. 
We first created randomized samples by keeping the original declinations of all 44 gold-type neutrinos while shuffling their R.A. coordinates. The arrival time of neutrinos in ICECAT-1 was preserved in our simulations since the time of neutrino arrivals is random and has negligible impact on our results. For each simulation, we obtained a count number $n$ of MIR-radio sources that matched with the neutrino events both spatially and temporally, following the methodology described in Section \ref{sample}.
%For each simulation, we obtain a count number $n$ of MIR-radio sources within the 90\% containment error radius of the neutrino events sample. 
The chance probability is calculated as the ratio between the number of the simulations that have $n \geq 1$ and the total number of simulations. From 30,000 Monte Carlo realizations, this approach results in a matching probability of $\sim 0.015$. An alternative estimation was performed by randomly redistributing the MIR-radio transients within SDSS survey footprints while keeping neutrino positions fixed. This yielded a chance probability of $\sim 0.011$ for coincident matches. 
In both cases, the chance probability for coincident matches from simulations is well consistent with the Poisson probability, corresponding to a significance for the association at a level of $\sim$99\%. 

\section{Analysis and Results}\label{results}
\subsection{Flaring Properties of J1513+3111}
We first examine the multi-wavelength electromagnetic radiation properties of J1513+3111 during the period of high-energy neutrinos arrival. Figure \ref{mir} (right) presents the WISE light curves up to 5 July 2023. 
It shows that the MIR emission peaked on 24 January 2017 (MJD=57778), preceding the neutrino arrival by 109 days. %, which is comparable to other 
In comparison to the two neutrino-associated TDEs from the MIRONG sample \cite{Jiang2023}, J1513+3111 also exhibits a larger variability amplitude ($\Delta$W2= 0.6) and a higher apparent luminosity (W2 = 12.09). %It is clear that the MIR light curve showed a fast rise and followed by a slow, monotonic decay to the post-flare level at $t\approx1986$ days since peak, which is consistent with the IR echo properties of the dusty TDE candidates \cite{Masterson2024}. 
Table \ref{flux} compiles all the radio data from both archival and our own follow-up observations. Details of data reduction and flux measurements can be found in Appendix.  

It can be seen that J1513+3111 was undetected in archival FIRST observations, with a 3$\sigma$ upper limit on the peak flux of 0.387 mJy. 
\src was later detected by VLASS epoch I at 3 GHz (on Oct 3, 2017), 142 days after the arrival of neutrino event. 
%across three epochs, there is no significant flux variability. Interestingly, 
%While \src displays only slight flux brightening at 3 GHz on Oct 9, 2020, 
There is a slight flux brightening in VLASS epoch II, followed by a decline in epoch III. 
More interestingly, the follow-up VLA observations between Nov 18, 2018 (MJD 58440) and Aug 15, 2021 (MJD 59441) 
%July 5, 2020 (MJD 59035) 
revealed a steady flux decline at $\sim$5.5 GHz, by a factor of $\sim$2.5, which can be seen in Figure \ref{SED_lc} (right).
Note that the flux variability is likely intrinsic, as the source is clearly detected at a high signal to noise ratio of $\simgt$20 with small flux errors. 
In addition, the flux variability cannot not be attributed to the resolution effect, 
as the first and last VLA observations have similar 
%beam sizes. 
spatial resolution. 
To further test the significance of variability in the light curve at 5.5 GHz differing from a constant (Figure \ref{SED_lc}, right), 
we performed $\chi^2$ analysis, and found that the radio flux is variable at a confidence level of $>$99.99\%, which is statistically significant. 

%from Nov 18, 2018 (MJD 58440) to July 5, 2020 (MJD 59035), 
%show a twofold decline in the radio flux between Nov 18, 2018 2018-11-18 (MJD 58440) and 2020-07-05 (MJD 59035). 
%J1513+3111 has a spectroscopically confirmed redshift of $ z = 0.07181$, which implies a luminosity distance $D_{L}\approx$ 324.23 Mpc assuming a flat cosmology with $\Omega_{\Lambda}$ = 0.7 and $H_0$ = 70 km s$^{-1}$ Mpc$^{-1}$. 
Due to the lack of known optical counterpart to the MIR flare in public surveys such as Asteroid Terrestrial-impact Last Alert System (ATLAS) \cite{Tonry2018}, the contemporaneous optical flaring emission of \src may be either obscured or intrinsically faint. 

In higher energy range, J1513+3111 was undetected in archival X-ray observations. Chandra observations on Dec 6, 2008, provided a $3 \sigma$ flux upper limit of $1.47\times 10^{-14}$ erg s$^{-1}$ cm$^{-2}$. 
Five years post neutrino detection, the Swift-XRT observations between March 15 and March 25 in 2022 
did not detect the source either, with a $3\sigma$ flux upper limit of $5.51\times 10 ^{-13}$ erg s$^{-1}$ cm$^{-2}$ in the 0.3–10 keV band. 
Since these X-ray observations did not cover the neutrino flaring period, it is not clear whether the X-ray 
emission has brightened or not. 
High-energy neutrino emission is usually accompanied by gamma-rays from electromagnetic cascade process. We also checked the \textsl{Fermi}-LAT data of J1513+3111 over the time interval that includes 300 days of observations before the arrival time of IC170514B. % on 2017 May 14. 
Following the standard analysis thread \cite{2022Univ....8..433P}, it is found that there is no significant gamma-ray emission in this time interval. We adjusted the time interval and chopped the gamma-ray light curve into a two-month time bin, with the duration of 600 days centered around neutrino arrival time. No gamma-ray emission was detected either. The upper limit energy flux at photon energies between 100 MeV and 800 GeV was estimated for a period of 300 days with photon power-law index $\Gamma = -2$, which is $1.05\times 10^{-12}$ erg cm$^{-2}$ s$^{-1}$ at a 95\% confidence level, consistent with results in \cite{Velzen2024}. 
%This is which is lower than the present gamma-ray observational limits by Fermi/LAT and HAWC.??
%\begin{table}
%    \centering
%    \caption{}
%    \begin{tabular}{cccccc}
%    \hline
%    \hline
%    Observatory & Project & Date & $\nu$ & $F_{\nu}$ & Beam Size\\
%    &     &     & (GHz) & (mJy/beam) & (arcsec$\times$arcsec)(deg)\\
%    \hline
%    FIRST & & 1994 June 5 & 1.4 &$\leq$0.387 & ... \\
%    \hline
%    GMRT & 37\_132 & 2019 Nov 11 & 1.25 & 0.632$\pm$0.018 & \\
%    \hline
%    VLA & 18B-086 & 2018 Nov 18 & 5.5 & 1.062$\pm$0.019 & 3.92$\times$3.39 (88.86) \\
%     & 20A-128 & 2020 July 5 & 5.5 & 0.536$\pm$0.0043 & 0.99$\times$0.91 (-34.43) \\
%     & & 2020 July 15 & 1.5 & 0.685$\pm$0.072 & 3.24$\times$3.01 (-52.85)\\
%     & 21A-146 & 2021 Aug 15 & 5.5 & 0.419$\pm$0.007 \\
%    \hline
%    EVN & ES091 & 2020 June 11 & 1.6 & 0.445$\pm$0.039 & 0.02$\times$0.01 (1.30)\\
%    & & 2020 June 16 & 4.9 & 0.380$\pm$0.030 & 0.00$\times$0.00 (11.29)\\
%    \hline
%    VLA & VLASS1 & 2017 Oct 3 & 3 & 0.555$\pm$0.038 & 3.67$\times$2.31 (74.62)\\
%    & VLASS2 & 2020 Oct 9 & 3 & 0.681$\pm$0.098 & 2.50$\times$2.36 (-79.46)\\
%    & VLASS3 & 2023 Jan 22 & 3 & 0.591$\pm$0.071 & 3.41$\times$2.34 (-78.42)\\
%    \hline
%    ASKAP & AS110 & 2021 Jan 1 & 1.4 & 1.327$\pm$0.031 & 19.50$\times$8.00 (11.82)\\
%    \hline
%    ASKAP & & 2019 Apr 21 & 0.9 & 0.888$\pm$0.051 & 25.08$\times$10.52 (18.81)\\
%    & & 2024 Jan 1 & 0.9 & 1.001$\pm$0.023 & 24.80$\times$11.20 (-4.46)\\
%    \hline
%    \end{tabular}
%    \label{flux}
%\end{table}

\begin{table*}[ht]
    \setlength{\abovecaptionskip}{10pt} % 调整 caption 上方间距
    \setlength{\belowcaptionskip}{10pt} % 调整 caption 下方间距
    \centering
    \begin{threeparttable}
    \caption {Summary of the radio observations of J1513+3111. The peak flux density observed by FIRST is 3$\sigma$ upper limit.} 
    \renewcommand{\arraystretch}{1.4}
    \setlength{\tabcolsep}{6pt} % 调整列间距
    \begin{tabular}{c|c c c c c c c} 
    \hline
    \hline
    Observatory & Project & Date & phase$^{\dagger}$ & $\nu$ & $F_{\nu}$$^{\star}$ & Beam Size\\
    &     &   & (days)  & (GHz) & (mJy/beam) & (arcsec$\times$arcsec)(deg)\\
    \hline
    FIRST & & 1994 June 5 & -8073 & 1.4 &$\leq$0.387 & ... \\
    \hline
    GMRT & 37\_132 & 2019 Nov 11 & 1217 & 1.25 & 0.632$\pm$0.036 & 3.49$\times$1.60 (78.72)\\
    \hline
    VLA & 18B-086 & 2018 Nov 18 & 859 & 5 & 1.158$\pm$0.033 & 3.45$\times$3.09 (89.10) \\
    & & 2018 Nov 18 & 859 & 6 & 0.937$\pm$0.016 & 2.76$\times$2.57 (-75.56) \\
     & 20A-128 & 2020 July 5 & 1454 & 5.5 & 0.536$\pm$0.027 & 0.99$\times$0.91 (-34.43) \\
     & & 2020 July 15 & 1464 & 1.5 & 0.685$\pm$0.080 & 3.24$\times$3.01 (-52.85)\\
     & 21A-146 & 2021 Aug 15 & 1860 & 5.5 & 0.419$\pm$0.022 & 3.75$\times$3.06 (81.96)\\
    \hline
    EVN & ES091 & 2020 June 11 & 1430 & 1.6 & 0.445$\pm$0.045 & 19.10$\times$8.40 (89)$^{*}$\\
    & & 2020 June 16 & 1435 & 4.9 & 0.380$\pm$0.036 & 3.61$\times$1.58 (6.9)$^{*}$\\
    \hline
    VLA & VLASS1 & 2017 Oct 3 & 448 & 3 & 0.555$\pm$0.047 & 3.67$\times$2.31 (74.62)\\
    & VLASS2 & 2020 Oct 9 & 1550 & 3 & 0.681$\pm$0.104 & 2.50$\times$2.36 (-79.46)\\
    & VLASS3 & 2023 Jan 22 & 2385 & 3 & 0.591$\pm$0.077 & 3.41$\times$2.34 (-78.42)\\
    \hline
    ASKAP & AS110 & 2021 Jan 1 & 1634 & 1.4 & 1.327$\pm$0.073 & 19.50$\times$8.00 (11.82)\\
    \hline
    ASKAP & & 2019 Apr 21 & 1013 & 0.9 & 0.888$\pm$0.068 & 25.08$\times$10.52 (18.81)\\
    & & 2024 Jan 1 & 2729 & 0.9 & 1.001$\pm$0.055 & 24.80$\times$11.20 (-4.46)\\
    \hline
    \end{tabular}
    \begin{tablenotes}
    \footnotesize
      \item$^{\dagger}$ The phases are measured in days relative to MIR discovery.
      \item$^{\star}$ The flux errors are calculated as the sum in quadrature of map rms and calibration uncertainty that is assumed to be of 5\% of the flux density.
      \item$^{*}$ The deconvolved beam sizes of EVN are given in units of milli-arcsecond, while the remaining observations are presented in clean beam sizes with arcsec units.
    \end{tablenotes}
    \end{threeparttable}
    \label{flux}
\end{table*}

\subsection{Radio SED and Morphology Analysis}\label{SED fitting}
Given the apparent radio flux variability, we investigated the temporal evolution of the radio spectral energy distribution (SED) with the synchrotron emission model in the context of an outflow-circumnuclear medium (CNM) interaction scenario. The shock generated therein accelerates the electrons and produces the transient radio synchrotron emission \cite{Giannios2011, Metzger2012}, as observed in \srcs. 
This model has been widely used to fit the radio emission from TDEs in which either a relativistic jet or a non-relativistic outflow has been launched 
\citep[e.g.,][]{Zauderer2011, Alexander2016, Horesh2021}. 
 With the measurements of synchrotron self-absorption frequency, $\nu_{a}$, and flux, $F_{\rm \nu, a}$ from the radio SED, we can estimate the size of the radio emitting region and its minimal energy assuming that the electron and magnetic field energy densities are in equipartition \cite{Barniol2013}.  
This is important for further exploring the relation between the radio-emitting outflow and the neutrino production site (Section \ref{discussion}).

\begin{figure*}
    \centering
    \includegraphics[scale =0.6]{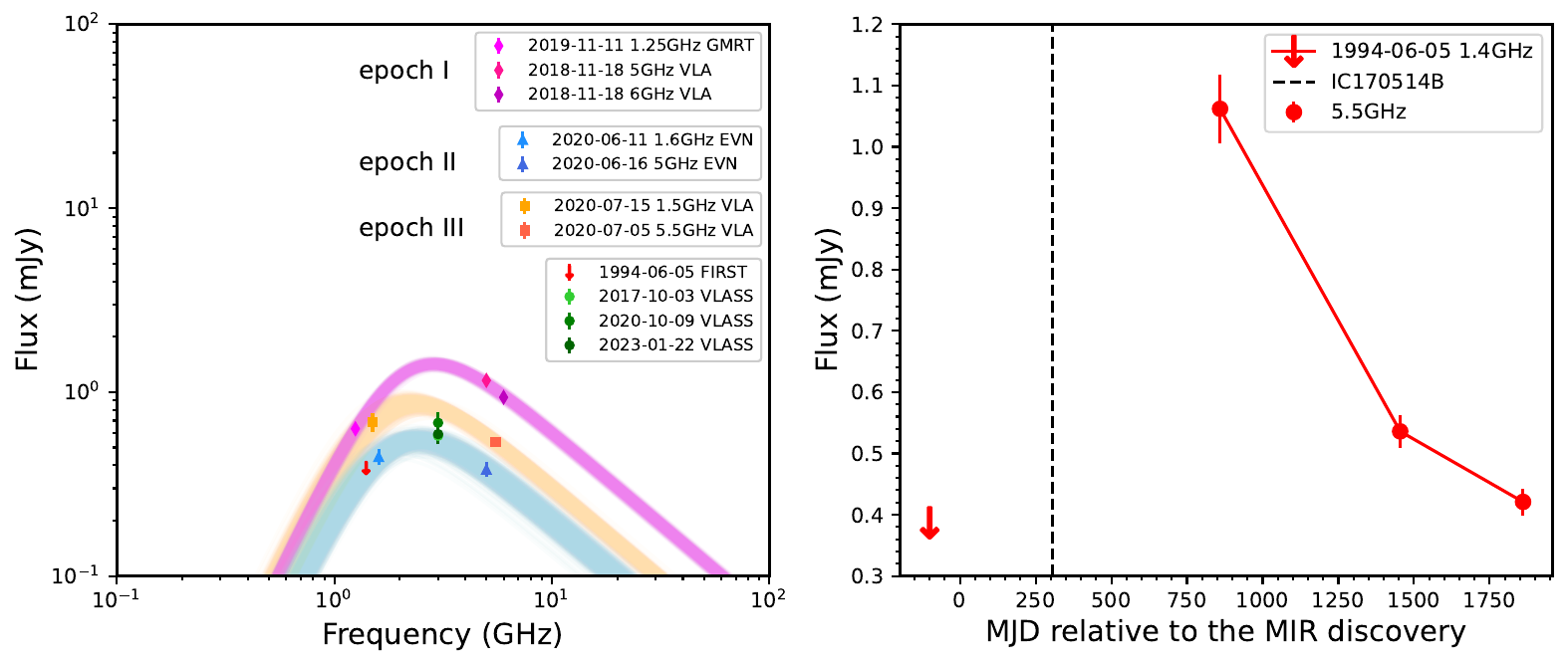}
    \caption{Left panel: Radio SED and its evolution over three epochs, using the data taken from VLA,  GMRT,and EVN observations. For the non-detections, the corresponding 3$\sigma$ upper limits on flux density are shown. Data for the three epochs are represented by magenta diamonds (epoch I), blue triangles (epoch II), and orange squares (epoch III). The color-coded lines represent the best fit to each SED from our MCMC modeling analysis, which are the model realizations on a basis of 500 random samples from the MCMC chains. Right panel: The radio flux density evolution of \src at 5.5 GHz. To better illustrate the transient nature of the radio emission, we show the 3$\sigma$ upper limit on the pre-flare radio flux at 1.4 GHz from FIRST survey (Table \ref{flux}), but with an arbitrary time relative to the MIR discovery.}
    
    \label{SED_lc}
\end{figure*}

As shown in Figure \ref{SED_lc} (left), the radio SED of J1513+3111 was constructed in 1.25 $\sim$ 6 GHz over three epochs covering an evolution period of $\sim$ 600 days. 
For the first epoch (Nov 2018), we divided the C-band VLA observations into two sub-band radio photometry centered at 5 GHz and 6 GHz, respectively, in order to better characterize the SED. 
In addition, due to the lack of low frequency observations in 2018, we took the data obtained from Giant Metrewave Radio Telescope (GMRT) in Nov 2019 to construct the radio SED in 1.25 $\sim$ 6 GHz. 
Although the VLA and GMRT observations were not performed quasi-simultaneously, the impact of potential intra-epoch flux variability is small, as the GMRT flux is consistent with that observed by VLA (on July 2020) at similar frequencies, indicating a slow evolution in the radio SED at low frequency ($\simlt$1 GHz) spanning $\sim$1 year.   
We tested that if the flux at 1.3 GHz is lower by a factor of 1.5 due to the actual SED evolution at earlier phases, the changes on the derived SED parameters are only at a level of 11\%-14\%. 
This would not affect the results from the further analysis of SED evolution. 
For the second and third epoch, the radio SED was constructed using the quasi-simultaneous observations at 1.5-5.5 GHz in June 2020 and July 2020, respectively (Table \ref{flux}). 

% centered at 5.5 GHz with 2 GHz aggregate bandwidth into two flux points. 
Note that the flux density observed by VLA in July 2020 is about 1.41 and 1.54 times the flux density at 1.6 GHz and 4.9 GHz obtained with the European VLBI Network (EVN), respectively. 
Rapid flux variability is very unlikely for such a short time-scale as the separation between the VLA and EVN observations is less than one month. 
Extended radio emission from nuclear outflow or jet is a natural explanation of the missing flux in the EVN images. 
%The extended radio emission is probably from a scale between 
Therefore, we excluded the EVN data from the following analysis of radio SED evolution, which  
displays a gradual shift to a lower peak flux density and frequency from the first to the third epoch. 
%However, the middle epoch, just one month prior to the final observation, unexpectedly breaks this pattern, presenting an intriguing observation puzzle. The deviation of fitting quasi-simultaneous EVN data may be due to the high resolution of the instrument (see Table \ref{flux}), which allows for the decomposition of the flux. 
%he EVN data are therefore excluded from SED analysis to avoid resolution-dependent flux bias. 

%Following the same approach outlined in \cite{Zhang2024}, the synchrotron emission spectrum is characterized by four parameters, namely F$_0$, $\nu_m$, $\nu_a$, and $p$, where F$_0$ is the flux normalization at $\nu_m$ (the synchrotron minimum frequency), $\nu_a$ is the synchrotron self-absorption frequency, and $p$ is the energy index of the power-law distribution of relativistic electrons. 

We fitted the SEDs with synchrotron emission models developed by \cite{Granot2002},  
%in the context of an outflow expanding into and shocking the surrounding medium, 
%specifically in the regime of $\nu_m \ll \nu_a$
assuming $\nu_m \ll \nu_a \ll \nu_c$, where $\nu_m$ is the characteristic synchrotron frequency of the emitting electrons with the least energy, $\nu_a$ 
is the self-absorption frequency and $\nu_c$ is the synchrotron cooling frequency.   %As in Goodwin et al. (2022), 
This is possible as our radio observations were performed at relatively late times ($\delta t >1000$ days), in which $\nu_m$ decreases more rapidly than $\nu_a$ due to the adiabatic evolution of the shock. 
The model SED is given by 
%\textbf{
\begin{equation}
\label{eqn:RadioSEDFunc}
\begin{split}
F_\nu = F_0\left[\left(\frac{\nu}{\nu_m}\right)^2e^{-s_{1}\left(\nu/\nu_{m}\right)^{2/3}}+\left(\frac{\nu}{\nu_m}\right)^{5/2}\right]\\
\times\left[1+\left(\frac{\nu}{\nu_a}\right)^{-s_{2}\left(\beta_1-\beta_2\right)}\right]^{-1/s_2}
\end{split}
\end{equation}
Here, $F_0$ is the flux normalization at $\nu_m$ (the synchrotron minimum frequency), and $p$ is the {\bf spectral index} of the power-law distribution of relativistic electrons. $\rm s$ represents the smoothing parameter, where $s_1=3.44p-1.41$ and $s_2=1.47-0.21p$. The spectral indices $\beta_1$ and $\beta_2$ correspond to the optically thick and thin regimes, respectively, with $\beta_1= 5/2$ and $\rm \beta_2 = -(p-1)/2$.
%where $\nu_m$ is the characteristic synchrotron frequency of the emitting electrons with the least energy and $\nu_a$ is the self-absorption frequency. 
The synchrotron emission spectrum is characterized by four parameters, F$_0$, $\nu_m$, $\nu_a$, and $p$.
%where F$_0$ is the flux normalization at $\nu_m$ (the synchrotron minimum frequency), and $p$ is the energy index of the power-law distribution of relativistic electrons. 
Owning to the limited data points especially at high frequencies ($>$10 GHz), the spectral index of the electron power-law distribution is fixed to $p=3$ (e.g. \cite{Cendes2021}). 
%We tested that if $p = 2.5$ is assumed, the minimal equipartition energy ($E_{\rm eq}$) and the equipartition radius ($R_{\rm eq}$) will decrease by $\sim$0.5 dex, which do not affect 
%the total energy of the system inferred from radio observations. 
%the outflow properties of radio-emitting region inferred.
Using a Markov chain Monte Carlo (MCMC) technique (python module emcee, \cite{Foreman-Mackey2013}), we can determine the best-fitting results of synchrotron model parameters and uncertainties. In Figure \ref{SED_lc} (left), we show the SED models which provide a reasonable fit to the data. Figure \ref{vp_Fp} shows the posterior distribution of the peak frequency $\nu_p$ and the peak flux density $F_{\nu,p}$, and the results are shown in Table \ref{parameters}.

%specifically the peak flux density $F_{\nu,p}$ and peak frequency $\nu_p$(Figure \ref{SED_lc}, left). 
We find that both $F_{\nu,p}$ and $\nu_p$ decrease steadily with time, from 1.43 mJy and 2.86 GHz to 0.87 mJy and 2.36 GHz, respectively.

With the inferred values of $F_{\nu,p}$ and $\nu_p$, we can further adopt an equipartition analysis to derive the radius of the radio emitting region ($R_{eq}$) and the nonthermal energy ($E_{eq}$) using the scaling relations outlined in \cite{Barniol2013}, 
%based on the energy minimization arguments which result in an equipartition between the electrons and the magnetic field. 
\begin{equation}\label{eq:E_eq}
\begin{split}
E_{\rm{eq}} &= E_e+E_B \\
&=1.3\times10^{48} \times 21.8^{-\frac{2(p+1)}{13+2p}}
(525^{p-1} \chi_{\rm{e}}^{2-p})^{\frac{11}{13+2p}}\\
&\quad \times (\frac{F_{\rm peak}}{1\,{\rm mJy}})^{\frac{14+3p}{13+2p}}
\left(\frac{d_L}{10^{28}\,\rm{cm}}\right)^{\frac{2(3p+14)}{13+2p}} 
\left(\frac{\nu_{\rm{peak}}}{10\,\rm{GHz}}\right)^{-1}  \\
&\quad \times (1+z)^{\frac{-27+5p}{13+2p}}
f_{\rm{A}}^{-\frac{3(p+1)}{13+2p}}
f_{\rm{V}}^{\frac{2(p+1)}{13+2p}} 
4^{\frac{11}{13+2p}} \quad \rm{erg},
\end{split}
\end{equation}
\begin{equation}\label{eq:R_eq}
\begin{split}
R_{\rm{eq}} &= 1\times10^{17}  (21.8 \times525^{p-1})^{\frac{1}{13+2p}}
    \chi_{\rm{e}}^{\frac{2-p}{13+2p}}
    (\frac{F_{\rm peak}}{1\,{\rm mJy}})^{\frac{6+p}{13+2p}}\\
&\quad \times \left(\frac{d_L}{10^{28}\,\rm{cm}}\right)^{\frac{2(p+6)}{13+2p}} 
    \left(\frac{\nu_{\rm{peak}}}{10\,\rm{GHz}}\right)^{-1}
    (1+z)^{-\frac{19+3p}{13+2p}} \\
&\quad \times f_{\rm{A}}^{-\frac{5+p}{13+2p}} f_{\rm{V}}^{-\frac{1}{13+2p}} 
    4^{\frac{1}{13+2p}} \quad \rm{cm},
\end{split}
\end{equation}
where $d_L$ is the luminosity distance ($\approx$$324$ Mpc) at $z=0.0718$, 
%is denoted by $d_L$ 
$f_A$ and $f_V$ are the area and volume filling  factors \cite{Barniol2013}, respectively.  
$E_e$ and $E_B$ are the energy in relativistic electrons and magnetic field, and the total energy is minimized with respect to $R$ at $R_{\rm eq}$, with $E_B=(6/11)E_e$ \cite{Barniol2013}. 
The factor $\chi_{\rm e}$ is expressed as $(\frac{p-2}{p-1}) \epsilon_{\rm e} \frac{m_{\rm p}}{m_{\rm e}}$, where $m_{\rm e}$ and $m_{\rm p}$ correspond to the mass of electron and the proton, respectively.
We can further correct the energy and radius for the system being out of equipartition using the following formulas:
\begin{equation}\label{eq:R}
\begin{aligned}
R = R_{\rm{eq}} \epsilon^{(1/17)},
\end{aligned}
\end{equation}
\begin{equation}\label{eq:E}
\begin{aligned}
E = E_{\rm{eq}}\left[(11/17)\epsilon^{(-6/17)} + (6/17)\epsilon^{(11/17)}\right],
\end{aligned}
\end{equation}
where $\epsilon=\frac{\epsilon_B}{\epsilon_e}\frac{11}{6}$. 
We assume that the fraction of the total energy carried by the relativistic electrons and the magnetic field is $\epsilon_e=0.1$ and $\epsilon_B=10^{-3}$, respectively. %, as suggested by observations of other TDEs \cite{Eftekhari2018}.   
In this case, the actual radius $R$ corresponding to the minimum energy deviates from $R_{\rm eq}$ by a multiplicative factor of 0.79 and the total nonthermal energy $E$ is greater than $E_{\rm eq}$ by a multiplicative factor of 2.68.

Following the procedures described in \cite{Goodwin2022}, we provide constraints for two different geometries, a spherical radiative zone with geometric factors $f_{\rm A}=1$ and $f_{\rm V}=4/3$, and a mildly collimated conical zone with a half-opening angle of $\phi=30^\circ$ and geometric factors $f_{\rm A}=0.13$ and $f_{\rm V}=0.178$.
%By further assuming the microphysical parameters $\epsilon_e$=0.1 and $\epsilon_B$=$10^{-3}$, which are the fractions of the total energy in electrons and magnetic field, respectively,  
%Both outflow models give a similar equilibrium radius ($R_{eq}\approx1\times10^{17}$ cm), with no remarkable time evolution in the 1$\sigma$ uncertainty. The derived kinetic energy of the outflow decreases by a factor $\sim 1.5$ (spherical) from $E_{eq}\approx6.61\times10^{49}$ erg to $E_{eq}\approx4.41\times10^{49}$ erg. 
%, which is higher than that of most other radio-emitting TDEs with non-relativistic outflows \cite{Cendes2024}. And that, under 
%Both outflow models give a similar
For the spherical outflow, we find the equipartition radius in epoch I and III is similar ($R_{eq}\approx1\times10^{17}$ cm), indicating no remarkable temporal evolution in the radio-emitting region spanning $\sim600$ days. 
If assuming a freely coasting outflow before our first detection of the radio emission ($\delta t=859$ days), we can place a lower limit on the outflow velocity of $v>0.04$c.  
%within the 1$\sigma$ uncertainty. 
The minimal energy of the outflow, 
that can explain the observed radio emission based on the equipartition analysis, decreases slightly 
%by a factor $\sim 1.5$ (spherical) 
from $E\approx4.4\times10^{49}$ erg to $E\approx2.9\times10^{49}$ erg over the same period. 
%Such a minimal energy is comparable to the specific binding energy of the most bound debris, if assuming a solar-type star was disrupted by a black hole with mass $M_{\rm BH}\approx10^6$\msun \cite{1988Natur.333..523R}. 
If assuming the mildly collimated conical outflow, 
%if assuming the conical outflow with a half-opening angle of 30$^\circ$, the kinetic 
the radius and outflow energy would be higher by a factor of $\sim$0.4 and 0.2 dex, but the evolution properties remain similar. 
%higher than that in the spherical case.

On the other hand, 
we tested that if $p = 2.5$ or $\epsilon_B=10^{-2}$ is assumed, the minimal equipartition energy ($E_{\rm eq}$) will decrease by $\sim$0.5 dex or $\sim$0.3 dex, while the equipartition radius ($R_{\rm eq}$) remains unchanged, which does not significantly affect the inferred properties of the radio-emitting region and outflow. 
Based on the assumed value of $\epsilon_e$, we estimate the shock energy to be $E_s = E_e / \epsilon_e \sim {\rm a~few} \times 10^{50}$ erg. The efficiency of electron acceleration in shocks is generally lower than that of protons, and theoretical studies indicate that the proton acceleration efficiency in a non-relativistic shock can reach up to 10-20\% at most \cite{2014ApJ...783...91C}. Therefore, a value of $\epsilon_e = 0.1$ may be largely overestimated and the shock energy may be higher than the above estimate. On the other hand, only a fraction of the outflow's kinetic energy is transferred to the shock, which is $\sim$10\% as found in simulations \cite{Hu2025}. Consequently, a conservative estimate suggests that the outflow kinetic energy 
%should be at least $10^{51}$ erg, or even much higher.  
could be at least $10^{51}$ erg, 
which is comparable to that inferred from simulations \cite{Hu2025}. 
%We conducted $p=2.5$ and $\epsilon_B=10^{-2}$ two sets of tests, as found in some other TDEs\cite{Cendes2022}. The equipartition radius $R_{eq}$ remains $\sim1\times10^{17}$ cm in two scenario, and the total kinetic energy $\rm E$ will decrease by $\sim$0.5 dex and 0.3dex, respectively, which do not affect the evolution properties inferred.

\begin{table*}[ht]
    \setlength{\abovecaptionskip}{10pt} % 调整 caption 上方间距
    \setlength{\belowcaptionskip}{10pt} % 调整 caption 下方间距
    \centering
    \begin{threeparttable}
    \caption {Fitted Parameters of the Synchrotron Model for J513+3111} 
    \renewcommand{\arraystretch}{1.4}
    \setlength{\tabcolsep}{6pt} % 调整列间距
    \begin{tabular}{c c c c c c c } 
    \hline
    \hline
& Epoch & $\delta t$ & $\nu_p$ & $F_{\nu,p}$ & $\log R_{eq}$ & $\log E $ \\
&  & $\rm (days)$ & $\rm (GHz)$ & $\rm (mJy)$ & ${\rm (cm)}$ & ${\rm (erg)}$ \\
\hline
Spherical & I & 859 & $2.86_{-0.07}^{+0.08}$ & $1.43_{-0.05}^{+0.05}$ & $16.99_{-0.01}^{+0.01}$ & $49.65_{-0.02}^{+0.02}$ \\ 
$f_A=1,f_V=4/3,\Gamma=1$ & III & 1454 & $2.36_{-0.15}^{+0.16}$ & $0.87_{-0.05}^{+0.05}$ & $16.98_{-0.03}^{+0.03}$ & $49.47_{-0.05}^{+0.05}$ \\ 
\hline
Conical & I & 859 & $2.86_{-0.08}^{+0.07}$ & $1.43_{-0.05}^{+0.05}$ & $17.41_{-0.01}^{+0.01}$ & $49.84_{-0.02}^{+0.02}$ \\ 
$f_A=0.13,f_V=0.178,\Gamma=1$ & III & 1454 & $2.36_{-0.15}^{+0.16}$ & $0.87_{-0.05}^{+0.05}$ & $17.40_{-0.03}^{+0.03}$ & $49.66_{-0.05}^{+0.05}$ \\ 
    \hline
    \end{tabular}
    \end{threeparttable}
    \label{parameters}
\end{table*}

\begin{figure}
    \centering
    \includegraphics[scale = 0.45]{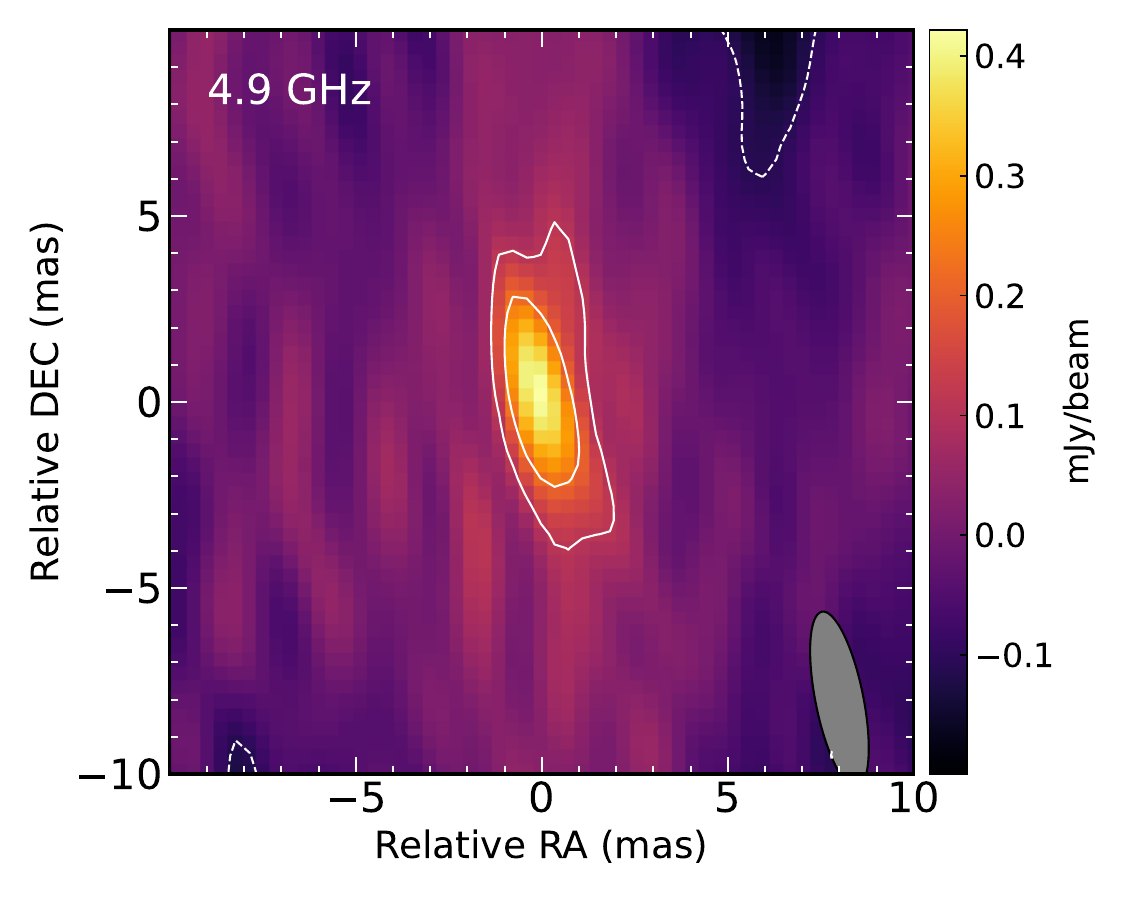}
    \caption{EVN image of J1513+3111 at 4.93 GHz, with a deconvolved size of 3.61 mas $\times$ 1.58 mas.  
    At the mas scale, \src remains compact with no significant extended emission. 
   % The contours are shown in (??)$\times$??$\sigma$ levels. 
    The gray filled ellipse in the right corner represents the shape of the beam.}
    \label{EVN}
\end{figure}

As shown in Figure \ref{EVN}, EVN detects a compact source in the final cleaned image (RA=15:13:45.760167, DEC=+31:11:25.057642), which has a deconvolved size of 3.61 mas $\times$ 1.58 mas at 4.93 GHz. To further investigate whether the source is resolved or not, we used the task Modelfit in DIFMAP to fit the radio emission component, but found no additional emission components in the residual map. Therefore, J1513+3111 remains compact and unresolved at the resolution of EVN observation, with an upper limit on its size of $<$ 2.1 pc, well consistent with the constraints on the radio source size from the radio spectral analysis. The brightness temperature of the compact radio emission can be estimated as 
\begin{equation}
    T_b=1.8\times10^9(1+z)\left(\frac{S_{\nu}}{1 {\rm mJy}}\right)\left(\frac{\nu}{1 {\rm GHz}}\right)^{-2}\left(\frac{\theta_1\theta_2}{1 {\rm mas}^2}\right)^{-1} {\rm K}
\end{equation}
where $S_\nu$ is the peak flux density in unit of mJy at the observing frequency $\nu$ in GHz, with $\theta_1$ and $\theta_2$ are the fitted FWHM of the major and minor axes of the Gaussian component in units of milliarcseconds. Using the deconvolved size ($\theta$) derived from the EVN image, we obtained a brightness temperature of $\sim 5.0 \times 10^6$ K. Note that the parsec-scale structure of J1513 is compact and not resolved with the EVN. Hence, only the upper limit on the source size can be constrained, and the brightness temperature should be considered as a lower limit.
The $T_b$ limit significantly exceeds the brightness temperature threshold of normal star-formation process (typically $\simlt10^5$ K), 
suggesting that the EVN component in \src is of non-thermal origin.

\section{Discussion}\label{discussion}
\subsection{Is J1513+3111 an Obscured TDE?}
%{\color{blue} the important evidence from Sun Luming...}
%The pre-flare SDSS spectrum 
The host galaxy of J1513+3111 can be classified as composite according to the line ratios of narrow emission lines in the pre-outburst SDSS spectrum \cite{Jiang2021b}. 
%The optical and radio centers are separated by only 57.68 milliarcseconds. This precise offset is consistent with the nuclear transient originating from the SMBH at the galactic center.
The transient position, as measured from our high-resolution EVN observations, shows a spatial offset of only 57.68 milliarcsecond from the host optical centroid. This corresponds to a physical offset of $\sim$79 pc at the redshift of \srcs, making it consistent with a nuclear origin. 
In addition, the X-ray luminosity before MIR outburst is less than $1.85\times10^{41}$ erg s$^{-1}$.  The combined optical and X-ray properties suggest that the galaxy exhibits little or weak AGN activity. The follow-up optical spectral observations after the MIR flare \cite{Wang2022} reveal transient broad H$\alpha$ and H$\beta$ emission lines that faded with time. They also reveal transient He \uppercase\expandafter{\romannumeral2} 4846 and Bowen N \uppercase\expandafter{\romannumeral3} broad emission lines and iron coronal narrow emission lines.
These features are commonly observed in TDEs \cite{Velzen2021, Hammerstein2023}. The MIR light curves in Figure \ref{mir} show a rise to peak over $\sim$ 197 days, then fading 0.5 mag over 729 days, settling into a plateau of 1257 days, followed by another decay to the pre-flare level at $t\approx1986$ days after the peak emission. The MIR flare in J1513+3111 can be classified as due to a TDE candidate with a probability of 86\% by \cite{2025arXiv250310053Y}, because the MIR color was initially blue and then turned red rapidly.
Indeed, the temporal evolution properties of MIR emission are comparable to other TDEs securely identified via multi-wavelength follow-up observations \citep[e.g.,][]{Stein2021, Huang2023}. 
The integrated energy radiated in MIR (host subtracted) is $\sim1.8\times10^{51}$ erg, which is also comparable to the energy observed in most TDEs \cite{Lu2018}. 
On the other hand, the black hole mass can be estimated to be $\sim5\times10^6$\msun \cite{Jiang2021}, which is typical to optical TDEs discovered by ZTF \cite{Yao2023}. 
Therefore, all these observational facts points to that J1513+3111 is likely an obscured TDE, and the possibility of an AGN outburst seems disfavored.

\subsection{Neutrino Production: $p\gamma$ Process}

In the context of a TDE, we discuss possible scenarios to account for the high-energy neutrino emission in \src if the association is genuine. 
\citet{Winter2023} constructed a detailed time-dependent quasi-isotropic neutrino production model ($p\gamma$ interaction) for three ZTF optical TDEs, including AT2019dsg, AT2019fdr, and AT2019aalc. 
Common to these three events, J1513+3111 also shows an elevated dust echo emission in infrared, which could be acted naturally as the target photons required for $p\gamma$ process. 
 
However, by fitting a blackbody model to the MIR emission in the peak for \srcs with the dust absorption coefficient considered, we found a MIR luminosity of $1.9\times10^{43}$\erg, the dust temperature of $\sim$1000 K or $kT\approx0.1$ eV, and radius of $\sim$0.13 pc or $4\times10^{17}$cm \citep{Jiang2021}. 
%Note that we 
Assuming the infrared photons are emitted from a spherical surface at a distance of 0.13 pc, the number density of these infrared photons is $n_{\rm IR}\sim 2\times 10^9$ cm$^{-3}$. The optical depth for photo-meson processes is $\tau_{p\gamma}=\sigma_{p\gamma}n_{\rm IR} \times 0.13~{\rm pc} \sim 0.1$, given that the $p\gamma$ cross-section $\sigma_{p\gamma} \sim 10^{-28}$ cm$^2$ \cite{2008PhRvD..78c4013K}. 
%{\bf While the $\tau_{p\gamma} \sim 0.1$ might be efficient for neutrino production, the energy mismatch still remains.}
%which is too low to efficient produce neutrino. 
In addition, in $p\gamma$ process, the energy of resulting neutrino is roughly $5\%$ that of the primary protons. For an observed neutrino, the energy of the target photons for the neutrino's primary proton is approximately $E_{\rm ph} \simeq 7~{\rm eV} (E_{\nu}/{\rm PeV})^{-1}$ (if considering head-on collisions, the target photon energy is 4 eV $(E_{\nu}/{\rm PeV})^{-1}$ ). 
The energies of protons undergoing $p\gamma$ reactions with these infrared photons are 1000 PeV. The resulting neutrino energies are predominantly in tens of PeV, which however, does not match the observed neutrino energy of sub PeV.  
It thus implies that $p\gamma$ interaction at the location of dust-emitting region may not be favored for J1513+3111.
However, such a scenario can be tested with future studies of the coincidence between MIR flares and neutrinos with even higher energies, such as observations by KM3NeT \cite{KM3NeT2025}. 
This will help to further constrain the physical process of proton acceleration and interactions %in the higher energy regime up to $\sim$1000 PeV. }
in the regime of $\sim$1000 PeV. 
 
On the other hand, for a sub-PeV neutrino, the corresponding energy of its primary proton is typically in the range of a few to several tens of PeV. The target photons for these protons are ultraviolet photons with energies of several tens of electron volts (hereafter, we adopt $E_{\rm{ph}} \geq 20$ eV). 
%For TDEs, the flux of UV photons is relatively high during the first few days, and then decreases over time. After one year of decay, we assumed that the UV luminosity has decreased to $10^{43}$ erg s$^{-1}$. 
For \src, we assume that the peak UV luminosity is comparable to that reprocessed in MIR by dust, which is $\approx10^{43}$ erg s$^{-1}$. 
The number density of UV photons is then $n_{\rm UV}\sim 5\times 10^{6}~{\rm cm^{-3}} [L({E_{\rm ph} \geq 20 ~{\rm eV}})/10^{43}~ {\rm erg~s^{-1}}] (r/0.13 {\rm pc})^{-2}$. The corresponding optical depth is $\tau_{p\gamma}=\sigma_{p\gamma}n_{\rm UV} \times 0.13~{\rm pc} \sim 10^{-4}$, 
%given that the $p\gamma$ cross-section $\sigma_{p\gamma} \sim 10^{-28}$ cm$^2$ \cite{2008PhRvD..78c4013K}, 
suggesting that $p\gamma$ interactions (with UV photons as target) at the location of the dust-emitting region are also inefficient. 
%In comparison, the number density of infrared photons is much higher, so the optical depth for those high-energy protons capable of interacting with them is expected to be greater. 
%However, as shown by the calculations above, 
The $p\gamma$ interaction with UV photons may become efficient at the scale of the inner accretion disk (on the order of $10^{-5}$ pc), since the optical depth of the interaction is inversely proportional to the distance from the reaction site to the black hole. 
Although it cannot be completely ruled out, this scenario is potentially unrelated to the infrared echo, and is difficult to reconcile with the significant time delay of the neutrino's arrival by $>100$ days (Section \ref{sample}). 

\subsection{Neutrino Production: $pp$ Process in the Context of Outflow-Cloud Interaction}
Given the detections of remarkable dust echo and broad emission lines in the optical spectra during the decline phase, gaseous clouds could exist in the circumnuclear region of \srcs. 
This motivates alternative scenarios that invoke $pp$ interaction in outflow-cloud model from TDEs to explain the production of neutrinos \cite{Wu2022}. 
Both observations and numerical simulations suggest that TDE can launch powerful outflow, 
which can propagate into the surrounding clouds if present, driving the bow shock to accelerate electrons and protons to relativistic energies efficiently. Shocked electrons in the magnetic field emit the synchrotron emission in radio band \cite{Mou2022}, while cosmic-ray proton interaction with medium in clouds accounts for the high-energy neutrino and gamma-ray emissions \cite{Wu2022}. 
This framework predicts the potential detections of flares across multiple messengers: neutrinos, transient radio emission, infrared echo, and gamma-rays (if not absorbed in the source). 
Following the outflow-cloud interaction model for AT2019dsg \cite{Wu2022}, we assume that for J1513+3111, a similar process occurs in the broad line region located at dozens of light-days ($\sim$0.01 pc) from the SMBH, 
as suggested by the optical reverberation-mapping observations of AGNs \cite[e.g.,][]{Bentz2013}. In this scenario, the high energy protons can be efficiently accelerated up to tens of PeV \cite{Wu2022}. We further assume that the total energy carried by the high-energy protons is $10^{51}$ erg, which is one order of magnitude higher than that of the relativistic electrons (Section \ref{SED fitting}), and their spectrum follows a power distribution, $dN/dE_p \propto E^{-p}_p$ ($E_p < 10$ PeV), with an index of $p=1.5$. Under these assumptions, the total amount of protons with $2 ~{\rm PeV} < E_p < 10 ~{\rm PeV}$ is $\simeq 8\times 10^{46}$, and these protons are expected to be responsible for producing sub-PeV neutrinos in the 0.1--0.5 PeV range. The inelastic collision cross-section for such high-energy protons is about 60 mb \cite{2006ApJ...647..692K}. If clouds possess a high column density exceeding $10^{25}$ cm$^{-2}$, such as the Compton-thick clouds invoked to explain the fast X-ray occurlation event in AGNs \cite{Miniutti2014, Kang2023},
or if strong and turbulent magnetic field  of about 1 Gauss \cite{Wu2022} threading the cloud traps protons, resulting in an effective column density 
increased to greater than $10^{25}$ cm$^{-2}$ (PeV proton's diffusion path within the cloud before escape can far exceed the cloud's size, thereby greatly boosting the effective column density), % due to repeated crossings of clouds by protons, 
the optical depth for these PeV protons is $\tau_{pp}=N_H \sigma_{pp} > 0.6$. 
This suggests that most of these protons are expected to undergo $pp$ collisions and produce neutrinos. 
Note that even if the clouds have a column density of a few times $10^{24}$ cm$^{-2}$, the optical depth for $pp$ collisions would be $\tau_{pp}\simgt 0.1$ and the neutrinos production can still be efficient.
For J1513+3111, the effective area of IceCube is $A_{\rm eff} \simeq 100\ {\rm m}^2$ in the energy range of 0.1--0.5 PeV \cite{2020PhRvL.124e1103A}, and thus the expected value of detected neutrinos is approximately $8\times 10^{46} /(4\pi d^2_L) \times A_{\rm eff} \simeq 0.01$. 
%Accordingly, we preliminarily infer that the $pp$ process may dominate the sub-PeV neutrino's production. 
Interestingly, by assuming that the outflow has a velocity of $v\sim$ 0.05c, as constrained from the modeling of the radio SED evolution (Section \ref{SED fitting}), we estimated that the outflow will travel by $t=d_c / v \approx240$ days to reach the clouds at $\sim$0.01 pc. 
%Such high-energy neutrino signal is expected to arrive with the time delay relative to the infrared flare. If the outflow reaches to the surrounding gaseous clouds at the distance of broad-line region ($\sim$0.01 pc) with an velocity of $\sim$0.05 c, the time delay can be estimate by 0.01pc/0.05c $>$ 200 days}, 
Considering the fact that the light traveling time for the dust-reprocessed IR emission is much shorter, typically tens of days, this suggests that the high-energy neutrino signal, if produced by the outflow-cloud interaction, will display a delay of $\sim$200 days relative to the time of IR outburst, 
%(with the light-traveling time of tens of days to respond to the accretion emission), }
which is comparable to what is observed for J1513+3111 (Figure ~\ref{mir}, right). 
Meanwhile, for these 2-10 PeV protons, the target photons required for the $p\gamma$ reactions must exceed $\sim 20$ eV. The number density of such target photons at the location of the broad line region is approximately $n_{\rm UV}\sim 1\times 10^{9}~{\rm cm^{-3}} [L({E_{\rm ph} \geq 20 ~{\rm eV}})/10^{43}~ {\rm erg~s^{-1}}] (r/0.01 {\rm pc})^{-2}$. 
%where we assume that the UV luminosity involved in the reaction is comparable to that of MIR. 
We estimate that the optical depth is $\tau_{p\gamma}\simeq n_{\rm UV} \sigma_{p\gamma}\times 0.01{\rm pc}\simeq 0.003$, suggesting that no more than 1\% of the PeV protons are likely to participate in the $p\gamma$ process, 
and the $pp$ process may dominate the sub-PeV neutrino's production. 

%\subsection{Radio–Neutrino Temporal Association}
In addition, pp collisions inevitably produce gamma-rays from $\pi^{0}$ decay. According to calculations by \cite{Wu2022}, gamma-ray photons with energies above 1 TeV are strongly absorbed by the extragalactic background before reaching the observer. For gamma-ray photons with energies below TeV, they are primarily produced by primary protons with energies below $\sim 10$ TeV. According to the assumed population of relativistic protons, the total energy of these protons is $3\times 10^{49}$ erg. Therefore, the total gamma-ray energy is a few times $10^{48}$ erg (taking $3\times 10^{48}$ erg for example). On the other hand, %since there are numerous clouds with a certain spatial distribution (rather than a single cloud), 
it can be assumed that most relativistic protons, trapped within the BLR clouds, undergo pp collisions over a timescale of one month. Under these assumptions, we derived that the gamma-ray luminosity is $1\times 10^{42}~{\rm erg~s^{-1}}$, and the corresponding flux is $1\times 10^{-13}~{\rm erg~cm^{-2}~s^{-1}}$, which is much lower than the Fermi-LAT limit. This is consistent with the non-detection of gamma rays.  

\subsection{Implications for Future Observations}

\begin{figure}
    \centering
    \includegraphics[scale = 0.6]{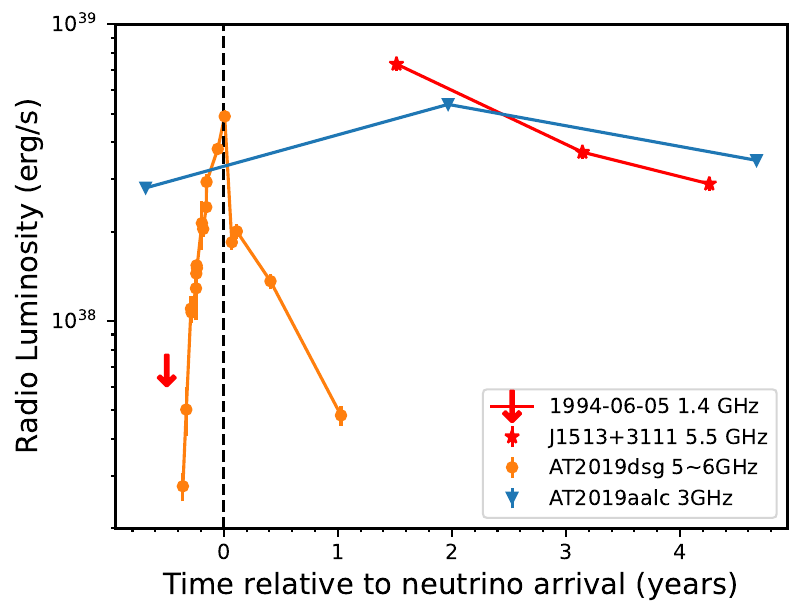}
    \caption{The 5.5 GHz radio luminosity evolution of \srcs. For non-detection, the corresponding 3$\sigma$ upper limit on luminosity is shown, with an arbitrary time relative to the neutrino arrival.
    %Also shown for comparison are the radio light curves of the two neutrino-coincident TDEs discovered by ZTF, AT2019dsg and AT2019aalc. 
    For AT2019dsg, we present only the data observed at $\sim$5-6 GHz. 
    For AT2019aalc, we show the 3 GHz luminosity evolution observed by VLASS, suggesting that the source experienced a radio brightening by a factor of 2 on a timescale of less than 1.5 years. 
    Note that there are no dedicated radio observations of AT2019aalc during the period of neutrino arrival, until ATCA observations that were performed about three years after the detection of neutrino \cite{Veres2024}. }
    %were observed at multiple frequencies, for simplicity we show only 5$\sim$6 GHz for each event. Due to the lack of multi-epoch radio observations, we show 3 GHz for AT2019aalc.} 
    \label{lum} 
\end{figure}

In Figure \ref{lum}, we plot the radio light curve of \src relative to the time of neutrino arrival, 
as well as a comparison to the two neutrino-associated ZTF TDEs with variable radio emission. 
%Except for AT2019dsg, the radio light curve is not well 
%Interestingly, the neutrino arrival time appears to coincide with the radio flux peak for AT2019dsg, 
Thanks to its relatively well sampled radio light curve (at $\approx$$5-6$ GHz), 
%it can be seen that the neutrino arrival time for AT2019dsg appears to coincide with the radio flux peak. 
AT2019dsg displays interesting coincidence between the radio flux peak and the neutrino arrival time, 
with little time delay ($\Delta t<$ 3 days). 
%consist with observation of good match between neutrino arrival time and radio peak flux time for AT2019dsg?? in Figure \ref{lum}. 
Owing to the absence of early radio observations, the peak of the radio light curve for J1513+3111 can not be constrained, but its luminosity should be much higher than that of AT2019dsg. For AT2019aalc, although the neutrino event appears to arrive in the rising phase of radio emission, as observed by VLASS over three epochs, the radio flare's properties (e.g., peak time and flux) remain unconstrained due to insufficient cadence in the radio observations. 
Due to the sparse sampling, we cannot use the radio light curve to further argue for the relation between the neutrino production site and the radio-emitting outflow in \srcs. 
%Notably, AT2019aalc exhibited a radio re-brightening four years after its discovery \cite{Veres2024}. Recent ATCA observations detected a prolonged radio flare with spectral inversion above 9 GHz, likely attributed to a newly launched outflow or jet component??.

In the future, the next-generation neutrino detectors with better angular resolution, such as the Giant Radio Array for Neutrino Detection (GRAND) \cite{Alvarez2020}, IceCube-Gen2 \cite{Aartsen2021}, and KM3NeT \cite{Adri2016}, will identify more high-energy neutrinos. %with better localizations. 
Meanwhile, the Square Kilometre Array (SKA), the world’s largest radio telescope operating across 50 MHz–15 GHz, will provide unprecedented angular resolution, survey speed, and sensitivity \cite{Braun2019}. 
%Its high-cadence monitoring will detect a growing population of radio transients with well-sampled light curves. 
In combination with the data from the optical time-domain surveys conducted by the Wide Field Survey Telescope \cite[WFST,][]{Wang2023} and the Large Synoptic Survey Telescope \cite[LSST,][]{Ivezic2019}, the IR observations by the Near-Earth Object Surveyor Mission  \cite{Mainzer2023}, 
and the X-ray surveys by Einstein Probe \cite{Yuan2025}, 
we will detect a growing population of TDEs with well-sampled multi-wavelength light curves covering the critical rise to peak phase down to a cadence of hours to days. 
These multi-wavelength datasets are important for a more uniform search for the spatial and temporal coincidence of TDEs with high-energy neutrinos, 
allowing for promising multi-messenger studies 
%of the radiation mechanisms of neutrinos and the acceleration mechanism of cosmic rays. 
to establish the TDEs as a source of neutrinos and pin down the dominant mechanisms producing neutrinos \cite{Winter2021,Winter2023,Wu2022,2023ApJ...954...17Z}. 
%involved in accelarating high energy cosmic rays, such as ??. 
. 

\section{Conclusion}
Our systematic search for radio transients in the MIRONG sample using the data from VLASS and dedicated VLA follow-up observations reveals that the TDE candidate J1513+3111 is spatially and temporally coincident with the track-type high-energy neutrino event IC170514B ($\sim$170 TeV), 
with a chance probability as low as $\sim$1.1\%. 
%Similar to the three previously reported neutrino-associated TDEs (AT2019dsg, AT2019fdr, and AT2019aalc), the neutrino detection time aligns with periods of elevated infrared and radio emissions for J1513+3111. These observational multi-messenger emissions, neutrino energy, and time delay, could be explained in the outflow-cloud interaction model. 
We analyze the radio flux and SED evolution properties, based on the data obtained from VLA and GMRT observations over two epochs spanning 605 days post MIR discovery. 
With a standard equipartition analysis, 
%in the context of an outflow interacting with CNM, 
we find that 
%a kinetic energy \textbf{of $E_{\rm K}>10^{50}$ erg} for outflow,  % (a conservative estimate), }
the size of radio emitting region remains little changed ($R_{\rm eq}\sim 10^{17} \ {\rm cm}$), while the shock energy (a few $10^{50}$ erg) decreases by a factor of $\sim1.5$. 
%\textbf{of $E_{\rm K}>10^{50}$ erg. 

%and radio-emitting region (potential neutrino production site) to be $\sim 10^{50} \ {\rm erg}$ and $\sim 10^{17} \ {\rm cm}$, respectively. 
%Using a standard equipartition analysis of the synchrotron spectral evolution spanning ?? days post mid-infrared discovery, we find a little evolution in the radio-emitting region, 
%significant  the outflow powering the radio emission is significantly decelerated, 
%while the minimum kinetic energy decreases from $E_{\rm K}\approx6.59\times10^{49}$erg to $E_{K}\approx4.41\times10^{49}$ erg. 
%The size of radio-emitting region can be further constrained by the 
High-resolution EVN imaging reveals a compact radio emission unresolved at a scale of $<$2.1 pc, 
with a brightness temperature of $T_b>5\times10^6$ K, suggesting that the transient radio emission 
is likely originating from the interaction between a TDE outflow and a dense circumnuclear medium. 
\src is the second TDE-neutrino association event (after AT2019dsg), for which the transient radio emission is robustly detected. 
%If the association is genuine, 
Our results highlight the role of TDE outflow-cloud interaction in producing the neutrino emission that is possibly related to the acceleration of protons through $pp$ collisions. 
We plan a systematic search for TDE-like transients within the VLASS catalog (Zhou et al., in preparation), aiming to uncover additional neutrino-linked radio-emitting TDEs and further test theoretical models.

%In summary, we have constructed a new sample to search for nuclear transients with radio and IR flares more efficiently. Through crossmatching, the discovery of J1513+3111 suggest that TDEs may be potentially associated with the origins of neutrinos. Among the current neutrino-associated TDEs, J1513+3111 is unique in exhibiting outflow characteristic, alongside AT2019dsg. Unfortunately, we are unable to determine whether the outflow emerged before or after the arrival of neutrino, it should be noted that the radio observations in early phase is significant, and it will help to understand the physics for the production of high-energy neutrinos. Furthermore, if outflows are commonly present in TDEs, and the SMBHs are surrounded by dense clouds, it could easily generate accelerated protons and high-energy neutrinos. Therefore, we will conduct the systematic search of TDE-like transients in radio band by VLASS catalog (L. Yang et al., in preparation) individually to find more neutrino-associated TDEs.
%or accreted with time-variable emissions

\section*{Acknowledgements}
We thank the anonymous reviewers for detailed and thorough comments that have improved the manuscript significantly.
%We thank the referee for very positive and constructive comments, which have improved the manuscript significantly.
We thank the staff of the GMRT, VLA and EVN, that made these observations possible. GMRT is run by the National Centre for Radio Astrophysics of the Tata Institute of Fundamental Research. 
The National Radio Astronomy Observatory is a facility of the National Science Foundation operated under cooperative agreement by Associated Universities, Inc.
The European VLBI Network is a joint facility of independent European, African, Asian, and North American radio astronomy institutes. Scientific results from data presented in this publication are derived from the following EVN project code: ES091. 
The Australian SKA Pathfinder is
part of the Australia Telescope National Facility, which is
managed by CSIRO. Operation of ASKAP is funded by the
Australian Government with support from the National
Collaborative Research Infrastructure Strategy. This paper
includes archived data obtained through the CSIRO ASKAP
Science Data Archive, CASDA (http://data.csiro.au). 
This research makes use of
data products from the Wide-field Infrared Survey Explorer, 
which is a joint project of the University of California, Los
Angeles, and the Jet Propulsion Laboratory/California Institute
of Technology, funded by the National Aeronautics and Space
Administration. 
GM thanks Kai Wang from HUST for helpful discussions. 
The work is supported by %the National SKA Program of China (2022SKA0130102), 
%the SKA Fast Radio Burst and High-Energy Transients Project (2022SKA0130102), 
the National Science Foundation of China (NSFC) through grant No. 12192220, 12192221, 11988101, 
and the National SKA Program of China (2022SKA0130102). 
X.S. acknowledges the science research grants from the China Manned Space Project with NO. CMS-CSST-2025-A07. %NO. CMSCSST-2021-A06. 
F.P. also acknowledge support from the Excellent Teacher Training Program of Anhui Province 2023 (YQZD2023007).  
G.M. is supported by the NSFC (No. 12473013). Y.C. thanks to the Center for Astronomical Mega-Science, Chinese Academy of Sciences, for the FAST distinguished young researcher fellowship (19-FAST-02). Y.C. also acknowledges the support from the National Natural Science Foundation of China (NSFC) under grant No. 12050410259 and the Ministry of Science and Technology (MOST) of China grant no. QNJ2021061003L.

\setcounter{table}{0}   %从0开始编号，显示出来表会A1开始编号
\renewcommand{\thetable}{B\arabic{table}}
\setcounter{figure}{0}
\renewcommand{\thefigure}{B\arabic{figure}}

\begin{figure*}[htbp!]
    \centering
    {  \includegraphics[scale = 0.6]{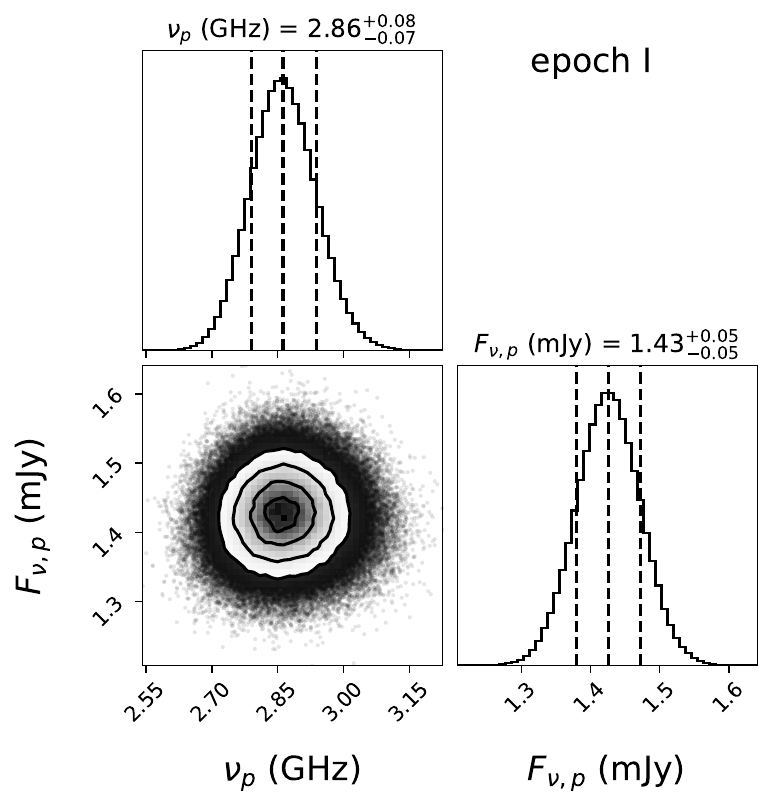}}
   {  \includegraphics[scale = 0.6]{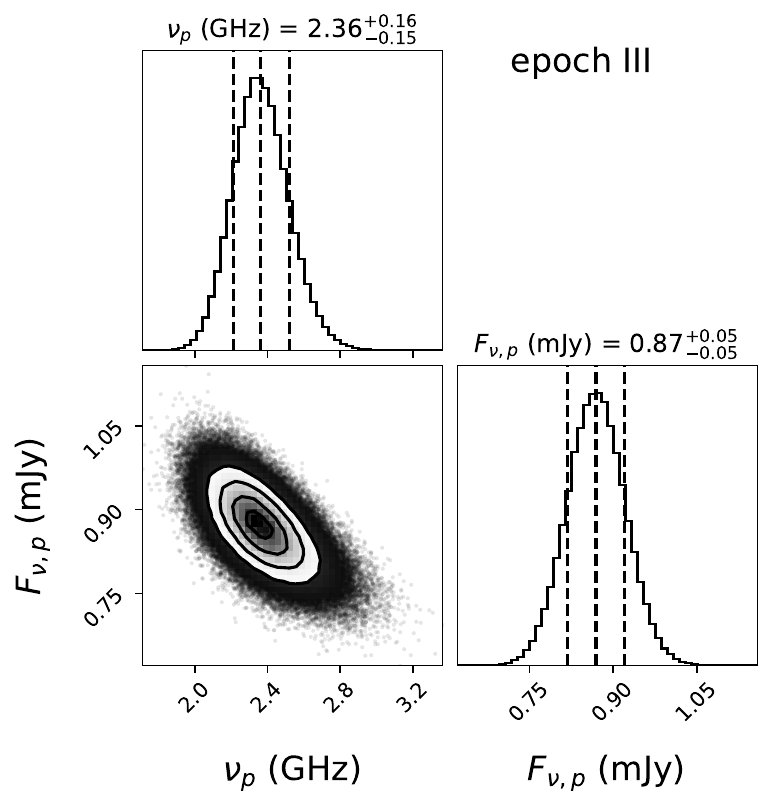}}
    \caption{Posterior distribution of the parameters spectral peak frequency $\nu_p$ and flux density at $F_{\nu,p}$, obtained by fitting the synchrotron spectrum to the observed radio SED. The dashed lines represent the 68\% quantile intervals.}
    \label{vp_Fp}
\end{figure*}

\setcounter{table}{0}   %从0开始编号，显示出来表会A1开始编号
\renewcommand{\thetable}{C\arabic{table}}
\setcounter{figure}{0}
\renewcommand{\thefigure}{C\arabic{figure}}

\begin{figure*}[htbp!]
    \centering
  %  { \includegraphics[scale = 0.6]{logT_logR.pdf}}
   {  \includegraphics[scale = 0.6]{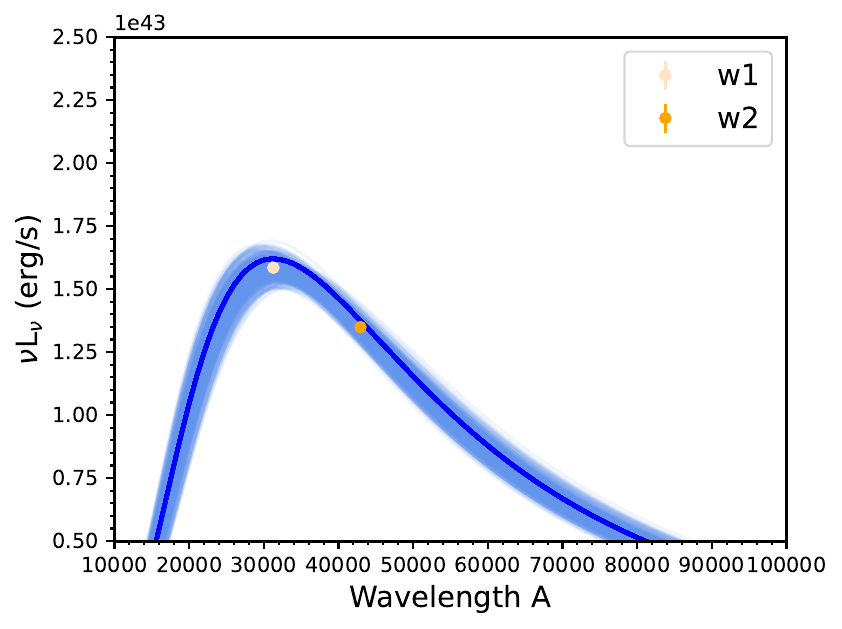}}
   { \includegraphics[scale = 0.6]{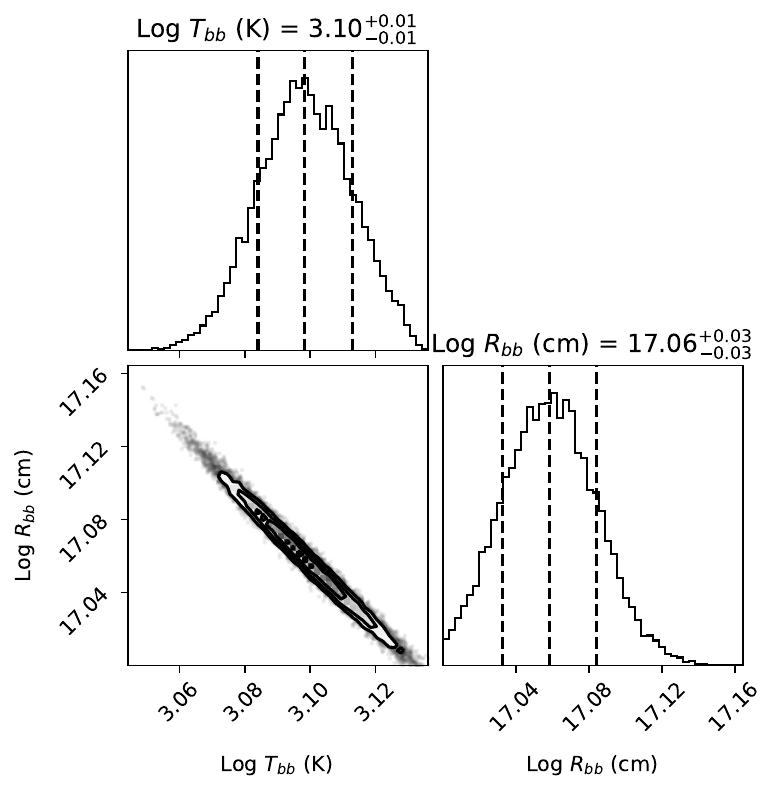}}
    \caption{ {\it Left panel}: The IR SED observed at the peak of MIR emission (the epoch of luminosity maximum at WISE W2-band). 
    The solid line represents the best-fit blackbody model, and the shaded region denotes the 1$\sigma$ error range of model realizations in MCMC fittings. {\it Right panel}: Posterior distribution of the parameters blackbody temperature $T_{bb}$ and blackbody radius $R_{bb}$, obtained by fitting the blackbody model to the data at the peak of MIR emission.   %at the luminosity maximum. 
    The dashed lines represent the 68\% quantile intervals. }
    \label{fitbb}
\end{figure*}

\appendix
\section{Radio Observations and Data Reduction}
\subsection{VLA}
To further study the origin and evolution of the radio emission, we conducted follow-up observations of J1513+3111 over three epochs using the VLA in C and L bands, centered at 5.5 GHz and 1.5 GHz, respectively. C-band observations were performed in C configuration on 2018 November 18 (project code: 18B-086), 2021 August 15 (project code: 21A-146), and the B configuration on 2020 July 05 (project code: 20A-128), while the L band was conducted on 2020 July 05. All the observations were phase-calibrated using the calibrator J1513+2338, and 3C 286 was used for bandpass and flux density calibration. The data were calibrated through the standard VLA calibration pipeline (version 2023.1.0.124) and analyzed with the Common Astronomy Software Applications (CASA, version 6.5.4 \cite{McMullin2007}). For the calibrated Measurement Set (MS), we applied additional flagging to channels affected by radio frequency interference (RFI) and split to sub-MS from groups of the spectral windows. The reduced data were imaged using the {\tt CLEAN} algorithm with Briggs weighting and ROBUST parameter of 0. We used the {\tt IMFIT} task in CASA to fit the radio emission component with a two-dimensional elliptical Gaussian model to determine the position, peak, and integrated flux density, as listed in the Table \ref{flux}.

\subsection{GMRT}
J1513+3111 was observed with the Giant Metrewave Radio Telescope (GMRT) at band 5 (central frequency of 1.25 GHz) on 2019 Nov 11 (project code: 37$\_$132). Flux calibration was conducted with 3C286, whereas the nearby bright source 1602+334 was also used to determine the complex gain solutions. The data from the GMRT observations were reduced using CASA (version 5.6.1) following standard threads and a pipeline adapted from the CAsa Pipeline-cum-Toolkit for Upgraded Giant Metrewave Radio Telescope data REduction \cite{Kale2021}. We began our reduction by flagging known bad channels, and the remaining RFI was flagged with the {\tt flagdata} task using the clip and tfcrop modes. We ran the task {\tt tclean} with the options of the multiscale multifrequency synthesis \cite{Rau2011} deconvolver, two Taylor terms (nterms = 2), and W-Projection \cite{Cornwell2008} to accurately model the wide bandwidth and the noncoplanar field of view of GMRT.

%J1513+3111 was detected in this observation, with a peak flux of 0.632 $\pm$ 0.018 mJy/beam. The uGMRT flux density measurements are shown in Table 1.

\subsection{EVN}

The EVN observations were carried out on 2020 June 11 and 2020 June 16 at central frequency of 1.66 GHz and 4.93 GHz, respectively (project code: ES091). Data from some individual telescopes were first temporarily stored at the station due to network problems and transferred to JIVE via the internet within two weeks after the end of the observations. The observations were performed in phase-reference mode by rapidly switching the telescopes between the target and a nearby bright calibrator (J1522+3144). The phase errors in the visibility data caused by the atmosphere can be solved by observing the bright phase-reference calibrator and applying the solutions to the target source. The phase-reference cycle was 5 min (3.5 min on the target and 1.5 min on the calibrator) during a total of 4.5h and 3.0h at 1.66 GHz and 4.93 GHz, respectively. We used the NRAO AIPS software to calibrate the amplitudes and phases of the visibility data, following the standard procedure from the AIPS Cookbook\footnote{\url{http://www.aips.nrao.edu/cook.html}}. The calibrated data were imported into the Caltech DIFMAP package \cite{Shepherd1997} for imaging and model-fitting.

\section{Fit the radio SED using the synchrotron model}
Figure \ref{vp_Fp} shows the posterior distribution of the parameters peak frequency $\nu_p$ and peak flux density at $F_{\nu,p}$, obtained by fitting the synchrotron spectrum to the observed radio SED (Figure \ref{SED_lc}, left). 

\section{Blackbody fits to the infrared SED}
We used the blackbody model to fit the SED at the peak of the MIR emission, which can constrain the blackbody temperature $T_{bb}$ and blackbody radius $R_{bb}$, as the blackbody emission has only two free parameters: 
\begin{equation}
    L_{\nu}=\pi B_{\nu}\left(T_{bb}\right)\times4\pi R_{bb}^2.
\end{equation}
Using the MCMC sampler $\tt emcee$ \cite{Foreman-Mackey2013}, we obtained the best-fitting results of $T_{bb}$, $R_{bb}$ and IR SED, which are presented in Figure \ref{fitbb}. For \src, we found the blackbody temperature is $T_{bb}$$\sim 1200$ K or $kT\approx0.1$eV and the blackbody radius is $R_{bb}$$\sim 0.04$ pc or $1\times10^{17}$ cm, which are consistent with the results in \cite{Jiang2021b}. 
Note that the real dust emission might not be a perfect blackbody, and the absorption coefficient should be considered. In this case, we found a blackbody temperature of $\sim$1000 K or $kT\approx0.1$eV, and radius of $\sim$0.13 pc or $4\times10^{17}$ cm, which have been used in our calculations of the $p\gamma$ interaction efficiency for IR photons.

\bibliographystyle{apsrev}
\bibliography{1513.bib}

\end{document}